\documentclass[aps,prb,twocolumn,groupedaddress,showpacs,floatfix]{revtex4-1}
\pdfoutput=1

\usepackage{graphicx}
\usepackage{dcolumn}
\usepackage{bm}
\usepackage{amsmath}

\begin{document}

\title{Unified microscopic approach to the interplay of pinned-Wigner-solid and liquid behavior 
of lowest-Landau-level states in the neighborhood of $\nu=1/3$}

\author{Constantine Yannouleas}
\email{Constantine.Yannouleas@physics.gatech.edu}
\author{Uzi Landman}
\email{Uzi.Landman@physics.gatech.edu}

\affiliation{School of Physics, Georgia Institute of Technology,
             Atlanta, Georgia 30332-0430}

\date{Submitted 15 May 2011; Phys. Rev. B {\bf 84}, 165327 (2011)}

\begin{abstract}
Recently observed microwave resonances in the spectrum of a two-dimensional electon gas under 
high magnetic fields in the neighborhood of the fractional filling $\nu=1/3$ were 
interpreted as signatures of a weakly pinned Wigner solid. Using the rotating-and-vibrating 
electron-molecule (RVEM) theory [Yannouleas and Landman, Phys. Rev. B {\bf 66}, 115315 (2002); 
Phys. Rev. A {\bf 81}, 023609 (2010)], in conjunction with exact diagonalization, a unified 
microscopic approach is developed for the interplay between liquid 
fractional-quantum-Hall-effect (FQHE) states and Wigner-solid states in the lowest Landau level
(LLL) in the neighborhood of $\nu=1/3$. In contrast to more traditional treatments, the RVEM 
theory utilizes a single class of variational wave functions for the description of both the FQHE 
liquid and Wigner-solid states, and their coexistence.

Liquid characteristics of the FQHE states are associated with the symmetry-conserving rotations
and vibrations of the electron molecule. The liquid characteristics, however, coexist with 
intrinsic correlations that are crystalline in nature, as revealed by the conditional 
probability distributions. Although the electron densities of the symmetry-conserving LLL 
states do not exhibit crystalline patterns, the intrinsic crystalline correlations are 
reflected in the emergence of cusp yrast states in the LLL spectra. These cusp states 
correspond to fractional fillings in the thermodynamic limit and are the only ones to provide 
the global ground states of the system. It is shown that away from the exact fractional 
fillings, weak pinning perturbations (due to weak disorder) may overcome the energy gaps 
between adjacent global states and generate pinned {\it broken symmetry\/} ground states as a 
superposition of symmetry-conserving LLL states with different total angular momenta. The 
electron densities of such mixed states (without good angular momentum quantum numbers) exhibit
oscillating patterns that correspond to molecular crystallites. These pinned Wigner crystallites 
represent finite-size precursors of the bulk Wigner-solid state. It is further shown that the 
emergence of these molecular crystallites is a consequence of the presence of RVEM components 
in the symmetry-conserving LLL states. In addition, it is shown that the RVEM approach accounts
for the Wigner-solid state in the neighborhood of $\nu=1$, which was also found in the 
experiments. Utilizing results for sizes in a wide range from $N=6$ to $N=29$ electrons, we address 
the extrapolation to the thermodynamic limit of the energetics of pinned Wigner crystallites, showing 
development of a crystal of enhanced stability due to contributions of quantum correlations. Furthermore, 
we address the size evolution of the crystal motifs (culminating in a hexagonal bulk two-dimensional 
Wigner lattice).
\end{abstract}


\maketitle

\section{Introduction}

In the early 1980's, the widely accepted theorical interpretation of the fractional quantum 
Hall effect (FQHE) phenomenon \cite{tsui82} was formulated around the antithesis 
between a new form of quantum fluid respresented by the celebrated Jastrow-Laughlin wave 
function \cite{laug83} and the pinned Hartree-Fock Wigner crystal (HFWC) described in the work of 
Fukuyama and Lee \cite{fuku78} and Maki and Zotos. \cite{maki83} In the above, the Wigner-crystal 
(WC) phase has been described \cite{fuku78,maki83} by broken-symmetry variational wave functions which
differ from those used for the quantum liquid. \cite{laug83}

The seminal paper by Laughlin \cite{laug83} pointed to two key aspects in favor of the 
``quantum-fluid'' interpretation: 

(i) The energy per particle of the HFWC varied smoothly with the filling factor $\nu$ in 
contrast to the experiment. However, Laughlin's wave function corresponded naturally 
to the major fractional fillings $[\nu=1/(2p+1)]$ that were observed experimentally [as a
result of the conservation of its total angular momentum $L=(2p+1)N(N-1)/2$].

(ii) The energy per particle (extrapolated to the thermodynamic limit $N\rightarrow \infty$) of the 
Laughlin-liquid state was substantially lower than that of the Wigner crystal at $\nu=1/3$. In the 
context of the major fractions, a crossover \cite{laug83} to a Wigner-crystal ground state \cite{tana89} 
was calculated to occur only for smaller fractional fillings; initially the onset
of Wigner-solid ground states had been estimated to occur for fillings $1/11 \leq \nu \leq
1/9$, while later studies \cite{lamg84} predicted crossover already for 
$1/7 \leq \nu \leq 1/5$. 

Based on the above studies, signatures of the Wigner crystal were expected to appear naturally 
in the range of smaller filling factors, and indeed over the last two decades experimental 
studies of the Wigner crystal in the lowest Landau level (LLL) seemed to validate
the above crossover prediction. \cite{andr88,li00,ye02,chen04}  Furthermore this crossover 
behavior between liquid (larger major fractions) and crystal (smaller major fractions) was 
also in agreement with the composite fermion (CF) approach for the 
liquid states, \cite{jain89,jainbook,jain00} including the modifications of the HF 
Wigner crystal referred to as composite fermion Wigner crystals. 
\cite{jainbook,yi98,nare01,chan05,chan06} (For an outline of the status of the CF theory for 
the Wigner crystal, see Sec.\ \ref{seccfc} below.)

In light of the above, the most recent observation \cite{zhu10.2} of experimental signatures 
associated with a pinned Wigner crystal in the immediate neighborhood of $\nu=1/3$ (as well as
\cite{zhu10.2,zhu10.1} in the neighborhood of $\nu=1$) represents a rather surprising 
development. In this paper, motivated by the above recent experimental observations, we further
develop the quantal theory of the rotating and vibrating electron-molecule (RVEM) description.
\cite{yl10} The RVEM incorporates liquid and crystalline correlations on an equal footing; it 
was introduced by us in previous publications \cite{yl10,yl02,yl03,yl04,yl07} and was shown 
\cite{yl10} to accurately describe the full LLL spectrum. In the RVEM theory, the description of
both liquid and Wigner-solid states is achieved within the framework of a single class of 
variational wave functions (see Refs.\ \onlinecite{yl02,yl10}). This allows us (see below) to discuss
the coexistence of FQHE liquid and Wigner-solid states. Namely, we will show that the 
application of this theory to the LLL states in the neighbohoods of $\nu=1/3$ and $\nu=1$ 
provides a {\it unified\/} microscopic interpretation (i.e., amenable to direct comparisons 
with exact solutions) pertaining to the emergence of both liquid-like and Wigner-solid behavior.
In addressing the emergence of the Wigner crystalline state and its coexistence with the FQHE liquid,
it is imperative that quantitative estimates of the energy difference between the liquid and solid
states be provided.

The plan of the paper is as follows:

Sec.\ \ref{secrvm} presents a brief outline of the RVEM trial functions and shows (in the
neighborhood of $\nu=1/3$) that the liquid characteristics of the LLL states coexist with 
intrinsic crystalline correlations revealed in the conditional probability distributions
[CPDs, see Eq.\ (\ref{defcpd})]. Further insights into the underlying physical reasons for this 
coexistence are given in Secs.\  \ref{secn6l45} and \ref{secn6l47}, where examples of 
quantitative analyses of the vibrational content (in terms of RVEM components) of the 
symmetry-conserving LLL states (obtained through exact diagonalization) are presented.

Sec.\ \ref{secpinn} describes (in the neighborhood of $\nu=1/3$) the effect of weak pinning 
(experimentally caused by weak disorder \cite{note31}) that generates broken-symmetry molecular (Wigner) 
crystallites manisfested in the electron density (ED) of the two-dimensional (2D) system. These 
broken-symmetry crystalline states result from the mixing of symmetry-conserving LLL states with 
different total angular momenta. 

Sec.\ \ref{secpinnu1} shows that the RVEM theory of liquid versus Wigner-crystallite behavior
can be extended to the neighborhood of $\nu=1$.

Our findings are not limited to the case of $N=6$ electrons (examined in detail in Secs.\
\ref{secrvm}, \ref{secpinn}, and \ref{secpinnu1}); they extend to larger sizes, as well.
Indeed Sec.\ \ref{seclarg} presents results for sizes in the range from $N=7$ to $N=29$ 
electrons. This section also addresses the extrapolation to the thermodynamic limit of the 
energetics (energy difference from the liquid FQHE state) of pinned Wigner crystallites, 
as well as the size evolution of the crystal motifs 
(culminating in a 2D hexagonal Wigner lattice for $1/N \rightarrow 0$).

Sec.\ \ref{seccfc} offers an outline of the status of composite-fermion literature regarding 
the challenging problem of a Wigner solid in the neighborhood of $\nu=1/3$.

A summary is given in Sec.\ \ref{secsumm}.

Appendix \ref{apprem} recapitulates the remaining analytic expressions needed to define the 
RVEM trial wave functions presented in Eq.\ (\ref{mol_trial_wf}). Furthermore, with the 
insights gained in this paper and the equivalence \cite{yl10} between the composite-fermion and
the RVEM theories, Appendix \ref{appcf} shows that intrinsic crystalline correlations are 
exhibited in the conditional probability distributions of the 
composite-fermion trial functions in the neighborhood of $\nu=1/3$. Moreover, Wigner 
crystallites (showing crystalline electron-density oscillations) are obtained via mixing of CF 
LLL states through the pinning process introduced in Sec.\ \ref{secpinn}. 

\section{Ro-vibrating electron molecule and the description of liquid-type behavior}
\label{secrvm}

As has been discussed earlier,\cite{jain95,lyl06} the many-body Hamiltonian (${\cal H}$) of an
assembly of $N$ electrons in the LLL is reduced to its two-body interaction (Coulombic) 
component, i.e.,
\begin{equation}
H^{\text{int}}_{\text{LLL}} = N \frac{\hbar \omega_c}{2} 
+ \sum_{i=1}^N \sum_{j > i}^N \frac{e^2}{\kappa r_{ij}},
\label{hlll}
\end{equation}
where $r_{ij}=|{\bf r}_i-{\bf r}_j|$, $\kappa$ is the dielectric constant, 
$\omega_c=eB/(m^* c)$ is the cyclotron frequency, $B$ is the applied magnetic field
perpendicular to the plane, and $m^*$ is the effective mass of the electron.

The neutralizing ionic background generates an overall external confinement.
\cite{taka86,yang02,wexl03,yang03,jola09,jola10} For high $B$, the external confinement contributes
an additional Hamiltonian term $H_{\text{LLL}}^{\text{con}}$ (see Sec.\ \ref{secpinn}) which 
influences only the total energies of the LLL states, but not their many-body structure. Its 
effects do not need to be considered in this section (such consideration will be postponed to
Sec.\ \ref{secpinn}).

The eigenstates of the Hamiltonian in Eq.\ (\ref{hlll}) have the property that they conserve 
the total angular momentum $L=\sum_{i=1}^N l_i$. This is instrumental in relating the 
precursor states of the finite system to the thermodynamic filling factors $\nu$ via the
relation \cite{laug83,girv83} 
\begin{equation}
\nu = L_0/L, 
\label{nu}
\end{equation}
with
\begin{equation}
L_0=N(N-1)/2.
\label{l0}
\end{equation}
For example, for a system of $N=6$ electrons, the lowest-energy state with a total angular 
momentum $L=45$ is the precursor that corresponds to the $\nu=1/3$ filling factor in the 
thermodynamic limit; for $N=7$ electrons, the corresponding state is the lowest-energy one with
$L=63$.

To determine the eigenstates and eigenenergies of the LLL Hamiltonian in Eq.\ (\ref{hlll}),
we employ two complementary approaches:
\begin{enumerate}

\item
The usual exact diagonalization (EXD) method which employs uncorrelated Slater determinants 
as the basis for the expansion of the many-body wave function. The number of Slater 
determinants in the expansion is referred to as the dimension of the EXD.
These Slater determinants are made out from the LLL single-particle orbitals 
\begin{equation}
u_l({\bf r}) = (2 \pi 2^l l!)^{-1/2} r^l e^{i l \phi} e^{-r^2/4},
\label{sporb}
\end{equation}
with lengths in units of the magnetic length $l_B=\sqrt{\hbar/m^* \omega_c}$. 

\begin{figure}[t]
\centering\includegraphics[width=7.4cm]{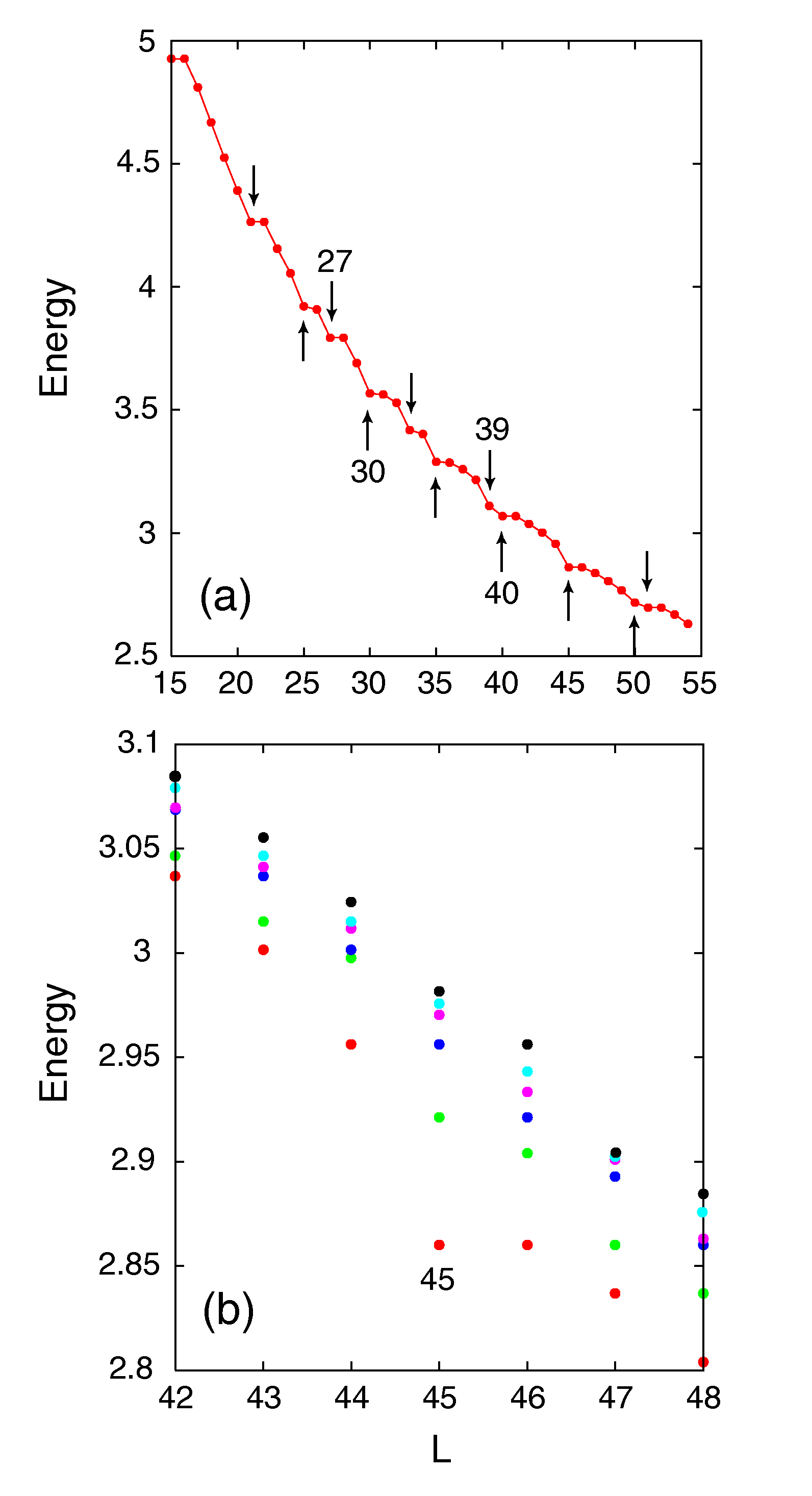}
\caption{(Color online) Exact-diagonalization energies for $N=6$ LLL electrons. Only the 
Hamiltonian term containing the two-body Coulomb interaction [see Eq.\ (\ref{hlll})] was 
considered. (a) LLL yrast states in the range $15 \leq L \leq 55$. The cusp states of the yrast 
line are marked by arrows. All the cusp states correspond to either a (1,5) (upward arrows) or 
to a (0,6) (downward arrows) Wigner-molecule ring configuration. \cite{yl03,yl07,note3} 
The cusp at $L=45$ occurs for both the $(1,5)$ and $(0,6)$ Wigner-molecule configurations. (b) 
The six lowest-in-energy states of the LLL spectrum in the immediate neighborhood of the 
magic angular momentum $L=45$ ($\nu=1/3$). Energies in units of $e^2/\kappa l_B$. 
The zero of the energy scale correspond to $N \hbar \omega_c/2$.}
\label{n6LLLspec}
\end{figure}

\item
The rotating-and-vibrating electron-molecule diagonalization (RVEM-diag) which was introduced 
in Ref.\ \onlinecite{yl10}. This method employs the technique of diagonalizing the LLL 
Hamiltonian in Eq.\ (\ref{hlll}) by expanding the many-body wave function in a correlated basis
constructed with the general ro-vibrational electron-molecule (RVEM) trial functions (within a 
normalization constant)
\begin{equation}
\Phi^{\text{RVEM}}_L =
\Phi^{\text{REM}}_{\cal L}(n_1,n_2,\ldots,n_r) Q[\Lambda] |0\rangle,
\label{mol_trial_wf}
\end{equation}
where 
\begin{equation}
Q[\Lambda] \equiv Q_{\lambda_1}^{m_1}Q_{\lambda_2}^{m_2} Q_{\lambda_3}^{m_3},
\label{qlam}
\end{equation}
with $\Lambda={\lambda_1}{m_1}+{\lambda_2}{m_2}+{\lambda_3}{m_3}$. The number of RVEM states
in the expansion is referred to as the dimension of the RVEM-diag; for a given $L$, this is much
smaller\cite{yl10} than the dimension used in the EXD.

The index REM stands for ``rotating electron molecule.'' [The terms Wigner molecule (WM) and
rotating Wigner molecule (RWM) are also often used; they are equivalent to electron molecule
(EM) and REM, respectively.] Here and in the following, $(n_1,n_2,\ldots,n_r)$ denotes an $N$ 
electron configuration consisting of concentric polygonal rings, with $n_1$ electrons in the 
innermost ring, $n_2$ electrons located in the second inner ring, ..., and $n_r$ electrons on 
the outermost ring; $N=\sum_{i=1}^r n_i$. The purely rotational (vibrationless) components 
$\Phi^{\text{REM}}_{\cal L}(n_1,n_2,\ldots,n_r)$ are associated with the cusp LLL states
(see Fig.\ \ref{n6LLLspec}) and have been described in detail in Refs.\ 
\onlinecite{yl02,yl03,yl07,yl10} (see brief description in Appendix \ref{apprem}).
The general RVEM wave function in Eq.\ (\ref{mol_trial_wf}) is a product that combines 
rotations with vibrational excitations, the latter being denoted by $Q_\lambda^m$, with 
$\lambda$ being an angular momentum; the superscript denotes raising to a power $m$. Both 
$\Phi^{\text{REM}}_{\cal L}$ and $Q[\Lambda]$ are homogeneous polynomials of the complex-number 
particle coordinates $z_1,z_2,\ldots,z_N$, of order ${\cal L}$ and $\Lambda$, respectively.
The total angular momentum $L={\cal L}+\Lambda$. $Q[\Lambda]$ is always symmetric in these
variables; $\Phi^{\text{REM}}_{\cal L}$ is antisymmetric (electrons are fermions). 
$|0\rangle$ is a product of Gaussians 
\begin{equation}
|0\rangle = \exp(-\sum_{i=1}^N z_i z_i^*/2),
\label{prod0}
\end{equation}
which is usually omitted from the notation.

The vibrational excitations $Q_\lambda$ are given \cite{note1,note2} by the symmetric 
polynomials:
\begin{equation}
Q_\lambda = \sum_{i=1}^N (z_i-z_c)^\lambda,
\label{ql} 
\end{equation}
where $z_c$ is the coordinate of the center of mass and $\lambda>1$ is an integer positive 
number. 
\end{enumerate} 

\begin{figure*}[t]
\centering\includegraphics[width=13.4cm]{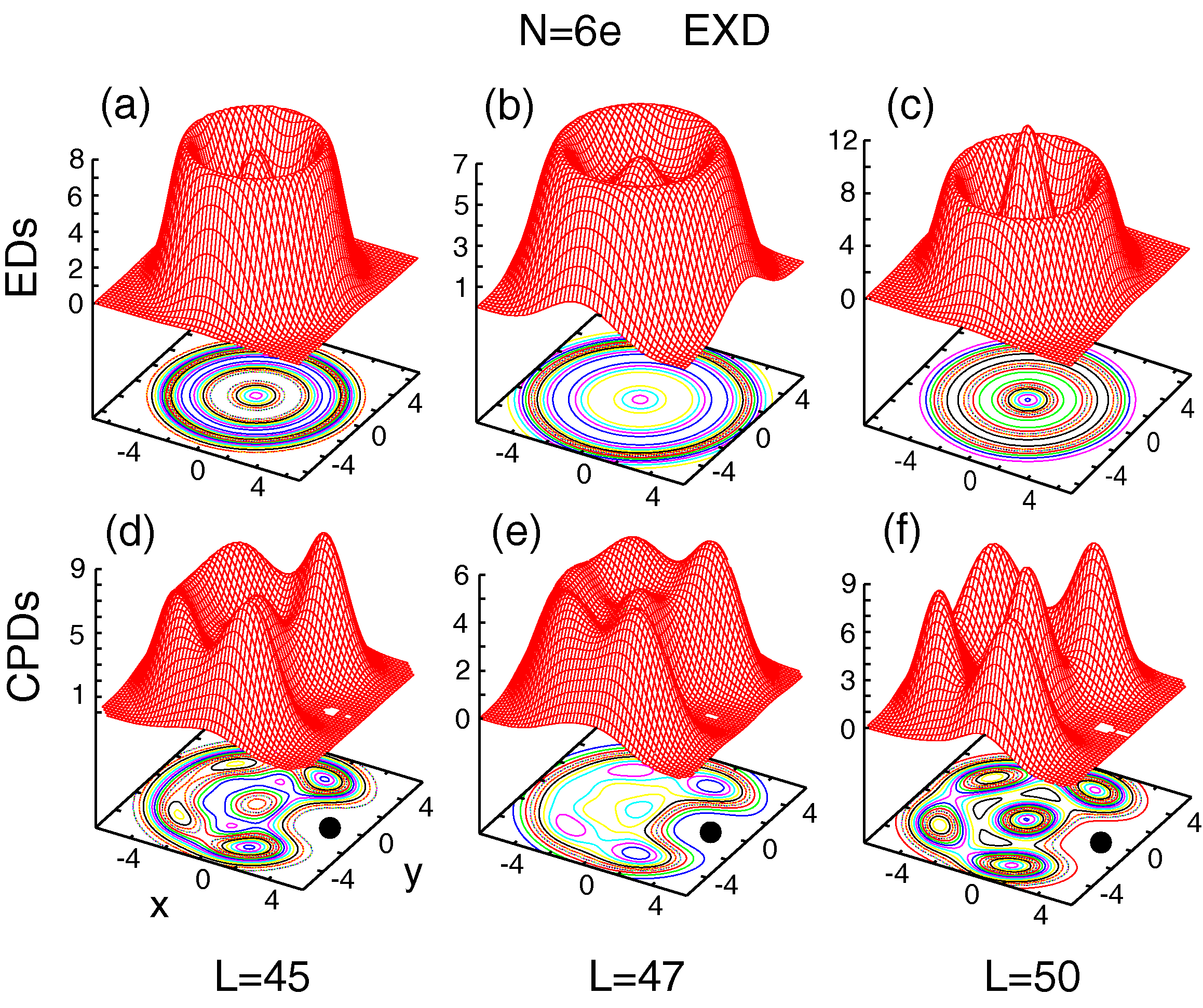}
\caption{(Color online) EXD electron densities (EDs, top row) and CPDs (bottom row) in the 
neighborhood of $\nu=1/3$ for $N=6$ LLL electrons, and for angular momenta (from left to right)
$L=45$ ($\nu=0.333$), 47, and 50. The solid dots denote the position 
of the fixed point. All three CPDs reveal the predominance (to various degrees) of the (1,5) 
molecular configuration. The units for the vertical axes in the CPD panels are arbitrary, but 
the same for all CPD frames here and throughout the paper. Lengths in units of $l_B$. The ED 
units are in $10^{-2} l_B^{-2}$. EDs are normalized to the number of particles, $N$.
}
\label{n6nu13exd}
\end{figure*}

As was shown in Ref.\ \onlinecite{yl10}, the RVEM-diag reproduces the EXD results to 
within arbitray precision. In this paper, we are not focussing on this numerical aspect. 
Rather we will use the RVEM-diag to analyze the extent that the vibrational degrees of 
freedom contribute to the exact wave functions in the neighborhood of $\nu=1/3$, in addition
to (and beyond) the vibrationless REM (the REM contibution by itself was studied in earlier 
publications, \cite{yl03,yl04} and naturally only for the exact fractional filling 
$\nu=1/3$). The importance of the vibrational components derives from the fact that the 
Laughlin trial functions (as shown in Ref.\ \onlinecite{yl10}), as well as the 
composite-fermion \cite{jain89,jainbook} ones (see Appendix \ref{appcf}), are expandable in the 
RVEM basis. This suggests an equivalent description of the ``liquid character'' \cite{jain00} 
of the LLL states within the framework of our quantal RVEM theory. Namely, states with a larger
weight of rotational-symmetry-preserving vibrational components exhibit enhanced liquid-like
character.

In the RVEM approach, the liquid state of the rotating and vibrating molecule is characterized by an 
azimuthally uniform liquid-like electron density (consistent with the fact that the RVEM wave functions
are eigenstates of the total angular momentum). Nevertheless crystalline correlations are
manifested in the CPDs, defined as 
\begin{equation}
P({\bf r},{\bf r}_0) =
\langle \Phi_L | \sum_{i \ne j} \delta({\bf r}_i - {\bf r}) \delta({\bf r}_j - {\bf r}_0)
| \Phi_L \rangle,
\label{defcpd} 
\end{equation}
where ${\bf r}_0$ is a fixed point in the intrinsic frame of reference of the rotating 
molecule. The CPD gives the probability of finding an electron at position ${\bf r}$ given that
another one is located at ${\bf r}_0$.

In this respect, as discussed in Sec.\ III of Ref.\ \onlinecite{yl04}, the rotating/vibrating 
electron molecule contrasts with the {\it nonrotating\/} (static) Wigner molecule familiar from 
unrestricted Hartree-Fock theories, which does not preserve the total angular momentum.
As a result the static Wigner molecule exhibits crystalline patterns in 
the electron density, and thus it is the proper finite analog of the bulk 
two-dimensional classical Hartree-Fock Wigner crystal (considered in the early paper of Maki 
and Zotos \cite{maki83}) and of its composite-fermion extension\cite{yi98} (composite-fermion 
Wigner crystal, CFWC). In the RVEM approach, behavior similar to a Wigner crystal is induced 
through pinning, as will be elaborated in Sec.\ \ref{secpinn}.

We return now to the description of LLL states in the neighborhood of $\nu=1/3$ having a good 
angular momentum $L$. In the context of precursor states in a finite system, Fig.\ 
\ref{n6nu13exd} displays EXD results (for electron densities, top row, and CPDs, bottom row) 
for three characteristic yrast states in this neighborhood. Specifically we consider $N=6$ 
electrons in the LLL with total angular momenta $L=45$ ($\nu=1/3=0.333$), 47, and
50. The EXD electron densities [Fig.\ \ref{n6nu13exd}(a-c)] are azimuthally 
uniform, in consonance with the quantum fluid picture of the LLL states. In contrast, for all 
three cases, the EXD calculated CPDs [Fig.\ \ref{n6nu13exd}(d-f)] exhibit crystalline
correlations reflecting the predominance of the (1,5) classical isomer \cite{beda94,kong03}
in the intrinsic frame of a rotating molecule. 

The crystalline-like EXD-calculated CPDs for cusp states (here for $L=45$ and $L=50$) have been
reported in many earlier studies (see, e.g., Refs.\ \onlinecite{yl03,yl04,tave03,bao95,seki96,
maks96}). Our EXD calculations (case of $N=6$ with $L=47$ in Fig.\ \ref{n6nu13exd} and results
for other $N$'s and $L$'s reported in Ref.\ \onlinecite{yl10}) demonstrate that the intrinsic 
crystalline correlations are present in all states comprising the LLL spectra.  

The degree of crystallinity (particle localization) in the CPDs of Fig.\ \ref{n6nu13exd} 
varies from one case to the other. This is due to the different weight of the vibrational 
modes $ Q_\lambda^m$ in comparison to that of the vibrationless REM component [see Eq.\ 
(\ref{mol_trial_wf})]. Naturally, a larger vibrational component in the EXD wave function 
results in reduced particle localization and in a relative enhancement of the liquid character 
of the LLL state. It is instructive to analyze the vibrational content of the LLL states in 
detail with the help of the RVEM theory. As illustrative examples, we consider below the cases 
of $N=6$ electrons with $L=45$ ($\nu=1/3$), which is an yrast cusp state) and $L=47$ 
(an yrast non-cusp state); see Fig.\ \ref{n6LLLspec}.

\begin{table}[t] 
\caption{\label{rvmdiagn6l45}%
Participation weights (sum of coefficients squared) of different subspaces to the RVEM-diag 
wave function for $N=6$ and $L=45$. The total dimension of the RVEM-diag space considered is 44 
(149 being the upper limit for the full TI subspace). The symbol 
$\Phi_{\cal L}^{\text{REM}}(n_1,n_2)Q[\Lambda]$ ($L={\cal L}+\Lambda$) denotes the subspace 
spanned by all the vibrations considered of the form 
$Q[\Lambda] \equiv Q_{\lambda_1}^{m_1}Q_{\lambda_2}^{m_2} Q_{\lambda_3}^{m_3}$ with
$\Lambda={\lambda_1}{m_1}+{\lambda_2}{m_2}+{\lambda_3}{m_3}$
}
\begin{ruledtabular}
\begin{tabular}{lrr}
${\text{RVEM subspace}}$ & ${\text{Dimension}}$ & ${\text{Weight}} $ \\ \hline 
$\Phi_{45}^{\text{REM}}(1,5)$        &  1  & 0.4477  \\
$\Phi_{40}^{\text{REM}}(1,5)Q[5]$    &  2  & 0.2344  \\
$\Phi_{35}^{\text{REM}}(1,5)Q[10]$   & 10  & 0.1490  \\
$\Phi_{30}^{\text{REM}}(1,5)Q[15]$   & 20  & 0.0912  \\
$\Phi_{45}^{\text{REM}}(0,6)$        &  1  & 0.0630  \\
$\Phi_{39}^{\text{REM}}(0,6)Q[6]$    &  5  & 0.0120  \\
$\Phi_{33}^{\text{REM}}(0,6)Q[12]$   &  5  & 0.0027  \\
\end{tabular}
\end{ruledtabular}
\end{table}

\subsection{The case $N=6$ with $L=45$ ($\nu=1/3=0.333)$}
\label{secn6l45}

For $N=6$ and $L=45$, the dimension of the EXD Hilbert space is 1206; that is the number of 
uncorrelated Slater determinants built out of the harmonic-oscillator states $u_l({\bf r})$
[see Eq.\ (\ref{sporb})]. The translationally invariant \cite{yl10} (TI) subspace spanned by 
the RVEM wave functions has a much smaller dimension of 149. Here we analyze RVEM-diag results 
(see TABLE \ref{rvmdiagn6l45}) for a smaller RVEM basis of dimension 44; this suffices to yield 
a many-body yrast state having an 0.990 overlap with the EXD wave function and an energy 
relative error of 0.141\% referenced to the EXD energy (i.e., an energy of 2.864187 
$e^2/\kappa l_B$ compared to the EXD energy of 2.860151 $e^2/\kappa l_B$).

\begin{figure}[t]
\centering\includegraphics[width=8.4cm]{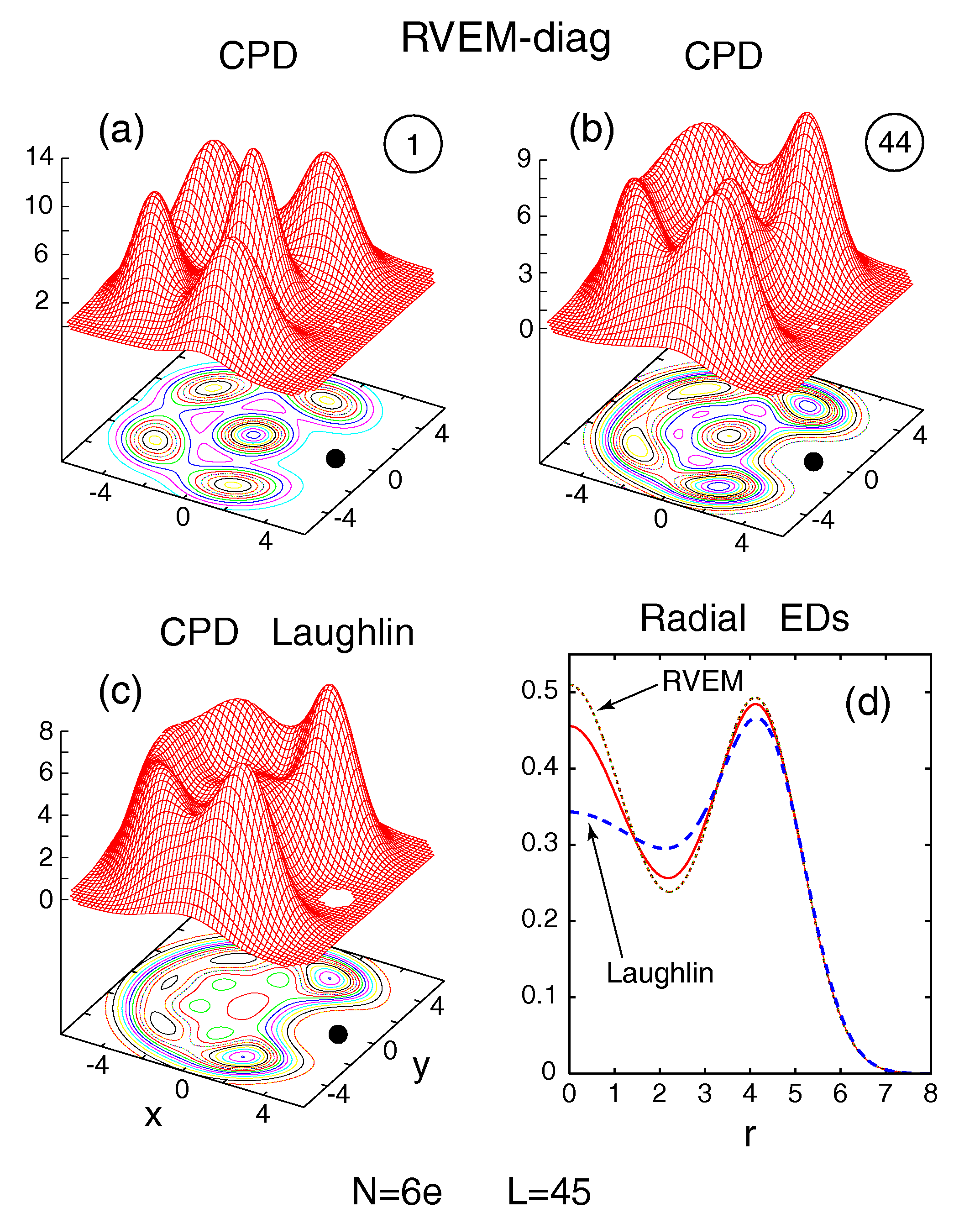}
\caption{(Color online) (a-b) RVEM-diag CPDs for the cusp yrast state with $N=6$ LLL electrons, 
and $L=45$ (corresponding to $\nu=1/3$). (a) CPD with only one RVEM state [namely, 
$\Phi_{45}^{\text{REM}}(1,5)$ with the largest participation, see TABLE \ref{rvmdiagn6l45}] 
included in the RVEM basis. (b) The CPD corresponding to the largest number of RVEM states 
considered in TABLE \ref{rvmdiagn6l45}. (c) The corresponding CPD for the Laughlin wave 
function. (d) The radial densities $\rho(r)$ for the EXD (solid line, online red), 
RVEM-diag (with 44 states, see TABLE \ref{rvmdiagn6l45}; dotted line, online brown), and 
Laughlin (long dashed line, online blue) wave functions. The solid dots in 
(a-c) denote the position of the fixed point. The CPD in (b) exhibits only minor differences 
from the EXD-calculated CPD in Fig.\ \ref{n6nu13exd}(d). The circled numbers in (a,b) denote 
the number of states included in the RVEM expansion. The units for the vertical axes in the 
CPD panels are arbitrary, but the same for all CPD frames here and throughout the paper. Lengths 
in units of $l_B$. The units of the vertical axis in (d) are $l_B^{-2}$. The radial densities are
normalized as $\int_0^\infty \rho(r) r dr=N$. }
\label{cpdn6l45rvmd}
\end{figure}

The state with $L=45$ is a cusp state. As a result the RVEM component with highest 
contribution is expected to have the form of a pure (vibrationless) 
$\Phi^{\text{REM}}_{45}(1,5)$, given that the (1,5) molecular configuration is predominant in
the corresponding EXD-calculated CPD [see Fig.\ \ref{n6nu13exd}(d)]. This expectation is confirmed by
the RVEM-diag results in TABLE \ref{rvmdiagn6l45}, where the participation weight (coefficient 
squared) of this $\Phi^{\text{REM}}_{45}(1,5)$ component is listed as 0.4477. We note that 
in total, including the vibrational components, the (1,5) isomer contributes the most with a 
participation weight of 0.9223, while the contribution of the (0,6) isomer is only 0.0777. 

In Figs.\ \ref{cpdn6l45rvmd}(a) and \ref{cpdn6l45rvmd}(b), the CPD of 
$\Phi_{45}^{\text{REM}}(1,5)$, which is the largest component in the RVEM-diag at $L=45$, is 
compared with the CPD associated with the RVEM diagonalization for the maximum expansion (44 
RVEM states) considered in TABLE \ref{rvmdiagn6l45}. The CPD of the pure vibrationless 
component [Fig.\ \ref{cpdn6l45rvmd}(a)] displays a (1,5) isomeric configuration with largest 
radial and azimuthal variations. The importance of the additional vibrational modes [containing
the $Q[\Lambda]$ factors] in bringing a close agreement with the EXD-calculated CPD is apparent [compare 
Fig.\ \ref{cpdn6l45rvmd}(b) with the EXD-calculated CPD in Fig.\ \ref{n6nu13exd}(d)].    

For $L=45$ ($\nu=1/3$ for $N=6$), it is natural to compare the behavior of RVEM-diag wave 
function (with 44 RVEM states, see TABLE \ref{rvmdiagn6l45}) with that of the Laughlin trial 
function, \cite{laug83,jainbook,yl04} in particular due to the fact that the corresponding 
energies differ only in the 4th decimal point. Indeed the Laughlin-state energy is 2.86440 
$e^2/\kappa l_B$ compared to the RVEM-diag energy of 2.864187 $e^2/\kappa l_B$;
this translates to a relative error of 0.148\% for the former compared to 0.141\% for the
latter.

To proceed in more depth with this comparison, we display in Fig.\ \ref{cpdn6l45rvmd}(c)
the CPD for the Laughlin state. It is apparent that the Laughlin CPD deviates from the CPD
associated with the EXD calculation [Fig.\ \ref{n6nu13exd}(d)] to a larger extent than the 
RVEM-diag one [Fig.\ \ref{cpdn6l45rvmd}(b)]; e.g., the central hump is significantly
attenuated in the Laughlin-state CPD, and this reinforces the impression of a ``liquid state.''
Furthermore, Fig.\ \ref{cpdn6l45rvmd}(d) compares the radial electron densities for the EXD, 
RVEM-diag, and Laughlin states. Again, the deviation between the EXD and RVEM-diag radial EDs 
is smaller than the deviation between the EXD and Laughlin radial EDs. This behavior is in 
agreement with the fact that the overlap between the EXD and RVEM-diag states is 0.990, while 
that between the EXD and the Laughlin state is smaller, \cite{tsip01,note4} i.e., 0.982.

\begin{table}[t] 
\caption{\label{rvmdiagn6l47}%
Participation weights (sum of coefficients squared) of different subspaces to the RVEM-diag 
wave function for $N=6$ and $L=47$. The total dimension of the RVEM-diag space considered is 78 
(180 being the upper limit for the full TI subspace). The symbol 
$\Phi_{\cal L}^{\text{REM}}(n_1,n_2)Q[\Lambda]$ ($L={\cal L}+\Lambda$) denotes the subspace 
spanned by all the vibrations considered of the form 
$Q[\Lambda] \equiv Q_{\lambda_1}^{m_1}Q_{\lambda_2}^{m_2} Q_{\lambda_3}^{m_3}$ with
$\Lambda={\lambda_1}{m_1}+{\lambda_2}{m_2}+{\lambda_3}{m_3}$.
}
\begin{ruledtabular}
\begin{tabular}{lrr}
${\text{RVEM subspace}}$ & ${\text{Dimension}}$ & ${\text{Weight}}$ \\ \hline 
$\Phi_{45}^{\text{REM}}(1,5)Q_2$     &  1  & 0.3549  \\
$\Phi_{40}^{\text{REM}}(1,5)Q[7]$    &  4  & 0.2283  \\
$\Phi_{35}^{\text{REM}}(1,5)Q[12]$   & 11  & 0.1485  \\
$\Phi_{30}^{\text{REM}}(1,5)Q[17]$   & 20  & 0.0400  \\
$\Phi_{45}^{\text{REM}}(0,6)Q_2$     &  1  & 0.1405  \\
$\Phi_{39}^{\text{REM}}(0,6)Q[8]$    &  8  & 0.0539  \\
$\Phi_{33}^{\text{REM}}(0,6)Q[14]$   & 22  & 0.0262  \\
$\Phi_{47}^{\text{REM}}(2,4)$        &  1  & 0.0002  \\
$\Phi_{43}^{\text{REM}}(2,4)Q[4]$    &  2  & 0.0012  \\
$\Phi_{39}^{\text{REM}}(2,4)Q[8]$    &  8  & 0.0059  \\
\end{tabular}
\end{ruledtabular}
\end{table}

\subsection{The case $N=6$ with $L=47$}
\label{secn6l47}

For $N=6$ and $L=47$, the dimension of the EXD Hilbert space is 1540. The translationally 
invariant \cite{yl10} (TI) subspace spanned by the RVEM wave functions has a much smaller
dimension of 180. Here we analyze RVEM-diag results (see TABLE \ref{rvmdiagn6l47}) for a 
smaller RVEM basis of dimension 78; this suffices to yield a many-body yrast state having an 
0.986 overlap with the EXD wave function and an energy relative error of 0.19\% referenced to 
the EXD energy.

The state with $L=47$ is not a cusp state [see Fig.\ \ref{n6LLLspec}(a)]. As a result the RVEM 
component with highest contribution is expected to have the form 
$\Phi^{\text{REM}}_{45}(1,5) Q_2 |0\rangle$, given that the (1,5) molecular configuration is 
predominant in the nearest $L=45$ cusp state (see Fig.\ \ref{n6nu13exd}), and corresponding to 
the fact that two units of angular momentum separate 47 from 45. This expectation is confirmed
by the RVEM-diag results in TABLE \ref{rvmdiagn6l47}, where the participation weight of this 
$Q_2$ component is listed to be 0.3549. We note that vibrations associated with the (1,5) 
isomer contribute the most with a (combined) participation weight of 0.7717, while those 
associated with the (0,6) isomer contribute only by 0.2206. The (2,4) isomer has a much smaller
contribution with a weight of 0.0073.

\begin{figure}[t]
\centering\includegraphics[width=8.4cm]{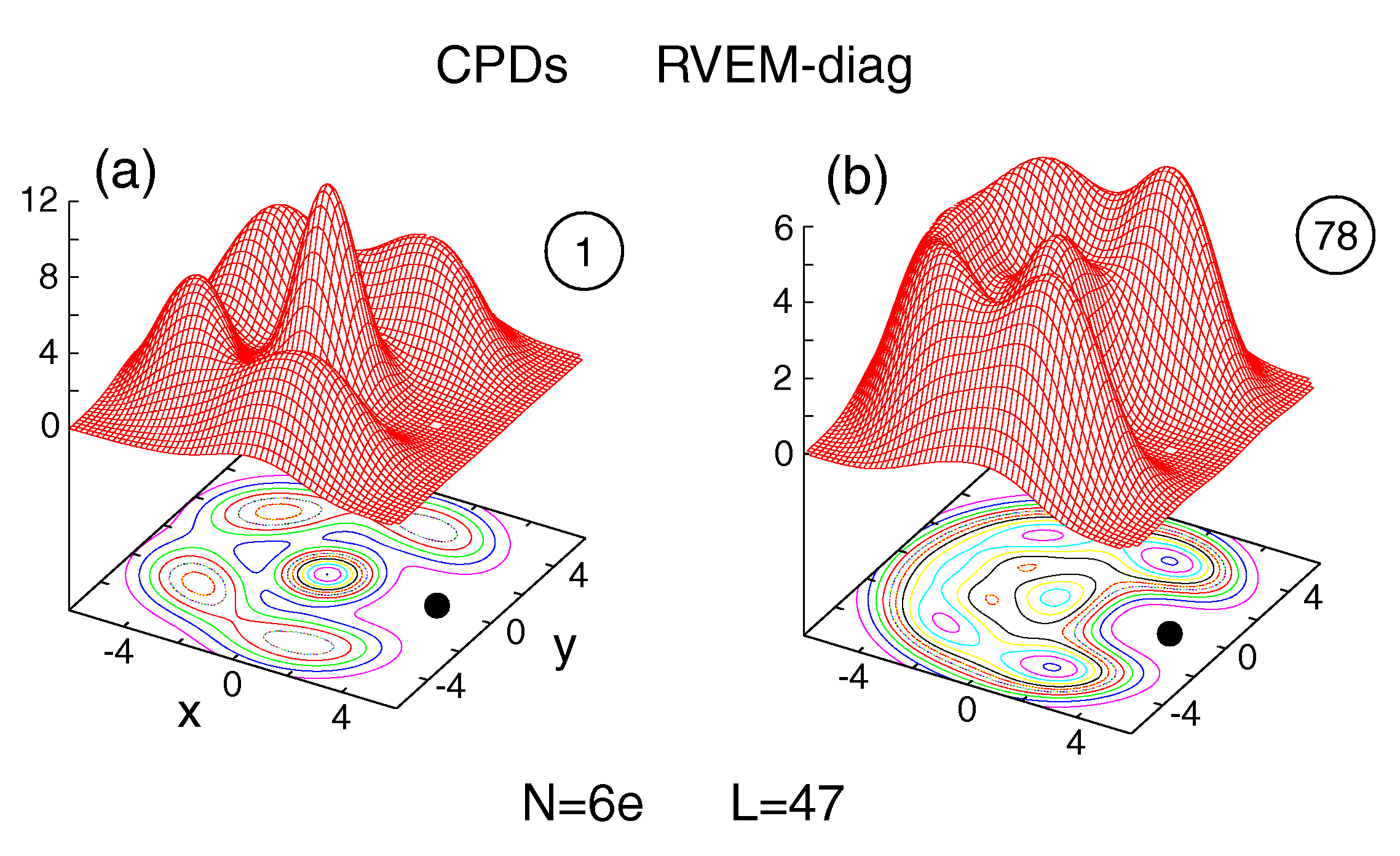}
\caption{(Color online) RVEM-diag CPDs for the non-cusp yrast state with $N=6$ and $L=47$.
(a) Only one RVEM state [namely, $\Phi_{45}^{\text{REM}}(1,5)Q_2$ with the largest
participation, see TABLE \ref{rvmdiagn6l47}] is included in the RVEM basis. (b) The CPD 
corresponding to the largest number of RVEM states considered in TABLE \ref{rvmdiagn6l47}.
The solid dots denote the position of the fixed point. The CPD in (b) exhibits only minor
differences from the EXD-calculated CPD in Fig.\ \ref{n6nu13exd}(e). The circled numbers denote the
number of states included in the RVEM expansion. The units for the vertical axes are arbitrary, 
but the same for all CPD frames here and throughout the paper. Lengths in units of $l_B$.
}
\label{cpdn6l47rvmd}
\end{figure}

Restating the above, we note that for $N=6$ and $L=47$ the vibrationless (1,5) component 
does not contribute to this LLL state. This is due to the fact that this state 
is a non-cusp yrast state. The component with the largest participation weight is 
$\Phi_{45}^{\text{REM}}(1,5)Q_2$; it corresponds to a dipolar ($\Lambda=2$) vibration of the 
largest component [i.e., $\Phi_{45}^{\text{REM}}(1,5)$] in the nearest cusp state with 
$L=45$.

In Fig.\ \ref{cpdn6l47rvmd}, the CPD of the largest component of the RVEM-diag calculated for 
$L=47$, that is $\Phi_{45}^{\text{REM}}(1,5)Q_2$, is compared with the CPD associated with the 
the wave function resulting from the RVEM diagonalization for the maximum expansion (78 RVEM 
states) considered in TABLE \ref{rvmdiagn6l47}. While the effect of the dipolar $Q_2$ vibration
in softening electron localization is visible when comparing to the pure REM CPD [compare Fig.\
\ref{cpdn6l47rvmd}(a) to Fig.\ \ref{cpdn6l45rvmd}(a)], the importance of the remaining 
additional vibrational modes in bringing a close agreement with the CPD obtained from the EXD
calculation is apparent [compare Fig.\ \ref{cpdn6l47rvmd}(b) to the EXD-calculated CPD in Fig.\ 
\ref{n6nu13exd}(e)].

\section{Pinned electron molecule and the description of crystal-type behavior}
\label{secpinn}

The experimentally observed rf or microwave resonances in the spectrum of a 2D
electon system under high $B$ have been interpreted \cite{zhu10.2,zhu10.1,li00,ye02,chen04} as
collective modes of a weakly pinned (due to disorder) Wigner-solid phase. Within the context 
of the LLL Hilbert space of a finite system, pinning can be described by a many-body 
Hamiltonian having the following two terms in addition to the Hamiltonian in Eq.\ (\ref{hlll}):
(i) impurity-type external potentials denoted by $V_{\text{imp}}$ and (ii) an overall 
confinement Hamiltonian term denoted by $H_{\text{con}}$. Namely,
\begin{equation}
{\cal H} = H^{\text{int}}_{\text{LLL}} + H_{\text{con}} + V_{\text{imp}}.
\label{mbh}
\end{equation}

The confinement Hamiltonian accounts for the neutralizing ionic background,
\cite{taka86,yang02,wexl03,yang03,jola09,jola10} and (for a smooth edge) it can be approximated as being 
harmonic
\begin{equation}
H_{\text{con}}=\sum_{i=1}^N 
\frac{1}{2m^*} \left( {\bf p}_i- \frac{e}{c}{\bf A}_i \right)^2 + 
\sum_{i=1}^N \frac{1}{2}m^*\omega_0^2 {\bf r}_i^2.
\label{hcon}
\end{equation} 
In Eq.\ (\ref{hcon}), ${\bf p}$ is the momentum of an electron and 
${\bf A} ( {\bf r} ) = (-By,Bx,0)/2$
is the vector potential.

In the presence of the confinement, the degeneracy of the single-particle orbitals within each
Landau level is lifted. In the high-magnetic field regime condidered in this paper 
($\omega_0 << \omega_c$), the harmonic-confinement part involves only the Darwin-Fock levels 
that form the LLL band, and it can be approximated \cite{jain95,lyl06} simply as
\begin{equation}
H_{\text{LLL}}^{\text{con}} = 
\hbar (\sqrt{\omega_0^2+\omega_c^2/4} - \omega_c/2) L.
\label{hconlll}
\end{equation}
Since $H_{\text{LLL}}^{\text{con}}$ is linear in the total angular momentum $L$, it influences 
only the total energies of the LLL states, but not their many-body structure [determined solely
by $H^{\text{int}}_{\text{LLL}}$, see Eq.\ (\ref{hlll})]. We note that the description of 
pinning below is independent of the precise value of $\omega_0$.

The first two terms in Eq.\ (\ref{mbh}) define the ``global'' Hamiltonian
\begin{equation}
H_{\text{glb}} = H^{\text{int}}_{\text{LLL}} + H_{\text{LLL}}^{\text{con}},
\label{hglb}
\end{equation}
which provides the global ground state of the system at a given $B$ (in the absence of 
disorder). 

\begin{figure}[t]
\centering\includegraphics[width=8.4cm]{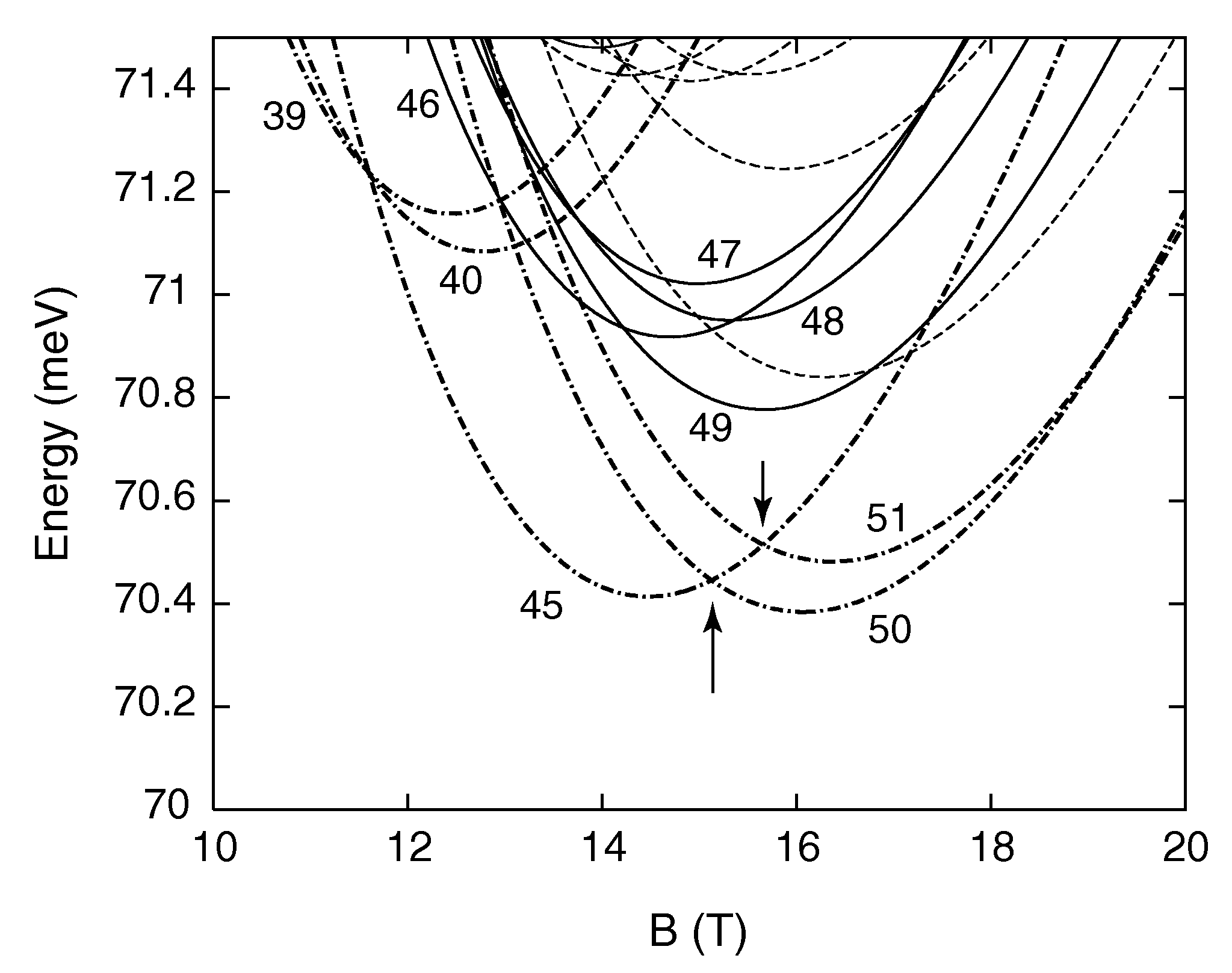}
\caption{The global spectrum as a function of the magnetic field in the 
neighborhood of $\nu=1/3$ for $N=6$ LLL electrons. The confinement was taken as 
$\hbar \omega_0 = 3.6$ meV. Note that all global ground states are yrast cusp states (see 
Fig.\ \ref{n6LLLspec}), but not all cusp states become global ground states. 
\cite{jain95,yl03,lyl06} The yrast cusp states with $L=39$ (0,6), 40 (1,5), 45 (1,5), 50 
(1,5), 51 (0,6) are portrayed by thick dashed-dotted lines. The $L=45$ curve relates to
$\nu=1/3$ in the thermodynamic limit. The numbers next to some curves (for yrast states 
only) denote the corresponding total angular momenta. The arrows highlight a couple of
curve crossings (for curves associated with cusp states). Remaining parameters: 
$\kappa=13.1$ and $m^*=0.067 m_e$, corresponding to GaAs. The topology (relative
position) of the curves is independent of the specific value for $\hbar \omega_0$ (see text).
}
\label{n6glbspec}
\end{figure}

An example of a global LLL spectrum (as a function of the applied magnetic field) 
corresponding to the Hamiltonian $H_{\text{glb}}$ is displayed in Fig.\ \ref{n6glbspec}.
Specifically, for $N=6$ electrons the global LLL spectrum is plotted in the neighborhood
of $\nu=1/3$ ($L=45$). It is seen that only cusp states (see Fig.\ \ref{n6LLLspec}) become
ground states [specifically for $L=40$, 45, and 50, associated with the (1,5) molecular
configuration]. The first excited states are separated from the ground states by relatively
large energy gaps and are composed of these (1,5) cusp states and those associated with the 
(0,6) molecular configuration (with $L=39$ and 51, see Fig.\ \ref{n6LLLspec}). \cite{note6}
The remaining LLL
states in Fig.\ \ref{n6LLLspec}, including the rest of the yrast states (e.g., with $L=41$,
42, 43, 44, 46, 47, 48, 49) become higher excitations in Fig.\ \ref{n6glbspec}.

The effect of the pinning perturbation term $V_{\text{imp}}$ in Eq.\ (\ref{mbh}) is to mix 
global states with good $L$ and produce a wave packet without a good total angular momentum. 
For a weak pinning case (small perturbation $V_{\text{imp}}$), it is apparent that 
$V_{\text{imp}}$ can efficiently mix only two global ground states in the neighbohood of their 
crossing points (denoted by arrows in Fig.\ \ref{n6glbspec}). Thus a weakly pinned state will
have in general the form
\begin{equation}
\Phi^{\text{PIN}}(L_1,L_2; \alpha,\beta)=\alpha \Phi_{L_1} + \beta e^{i\theta} \Phi_{L_2},
\label{pinnwf}
\end{equation}
where $L_1$ and $L_2$ are the magic angular momenta of the global ground states and 
$\alpha^2+\beta^2=1$. The phase $\theta$ determines the orientation of the pinned state; we 
mention it here for the sake of generality and completeness, but it is not an essential 
parameter for the rest of the paper.

It is apparent that the total angular momentum of the wave packet state
in Eq.\ (\ref{pinnwf}) is not a good quantum number and is given as the average value
\begin{equation}
\bar{L}=\alpha^2 L_1 + \beta^2 L_2. 
\end{equation}
Likewise, the energy of the pinned state is given as the average of the energies $E_1$ and $E_2$ of
the superimposed states
\begin{equation}
E^{\text{PIN}} (L_1, L_2, \alpha, \beta) = \alpha^2 E_1 + \beta^2 E_2.
\label{enpin}
\end{equation}

\begin{figure}[t]
\centering\includegraphics[width=8.4cm]{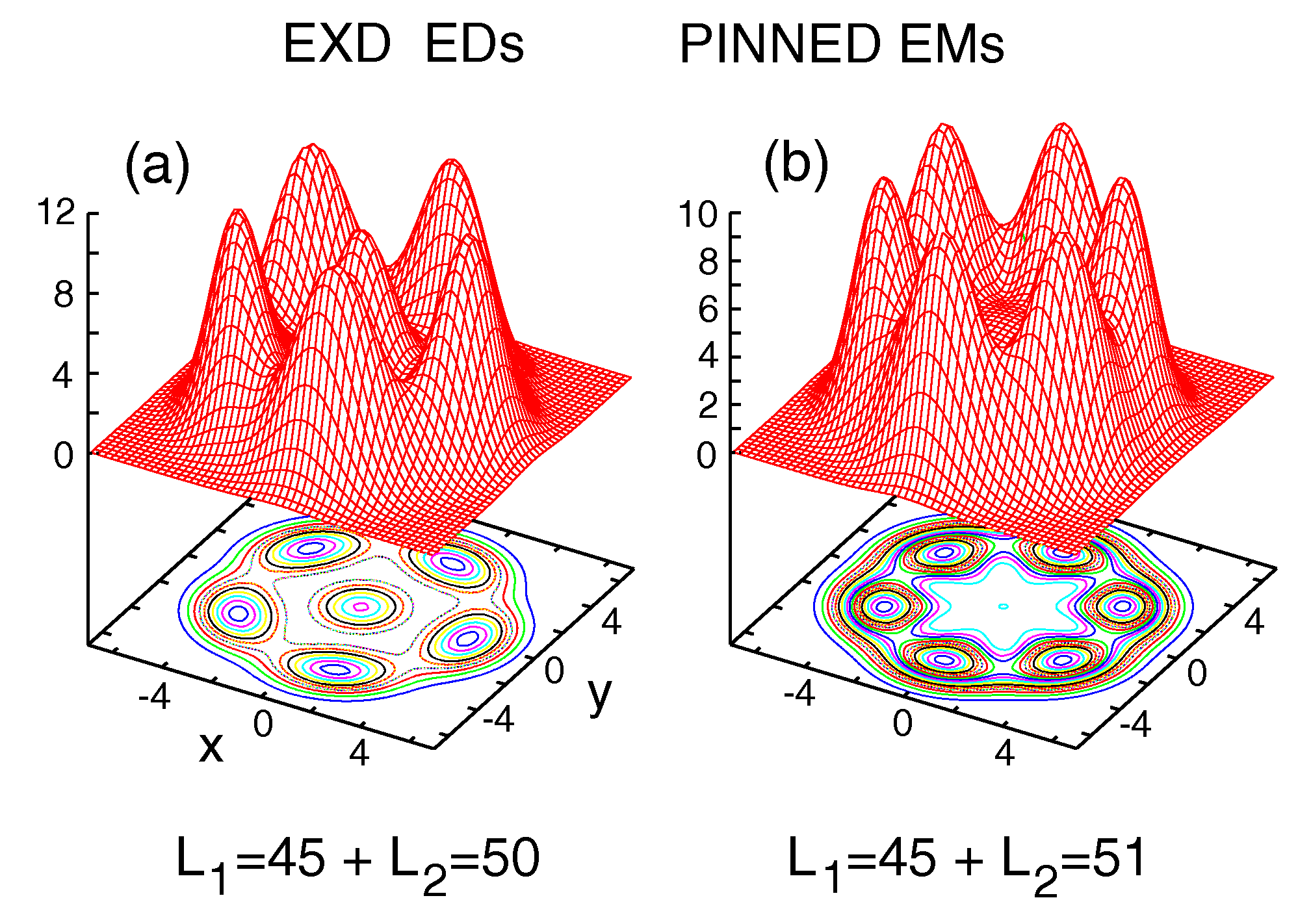}
\caption{(Color online) Electron densities for pinned [see Eq.\ (\ref{pinnwf})] LLL states in
the neighborhood of $\nu=1/3$ for $N=6$ electrons. EXD states have been used for both 
$\Phi_{L_1}$ and $\Phi_{L_2}$. (a) $L_1=45$ and $L_2=50$ ($|L_1-L_2|=5$). (b) $L_1=45$ and 
$L_2=51$ ($|L_1-L_2|=6$). The formation of a pinned (nonrotating) EM representing a (1,5) 
molecular configuration in (a) and a (0,6) molecular configuration in (b) is transparent. 
$\alpha=\beta=1/\sqrt{2}$. Note that all six humps of localized electrons are visible in the
electron densities of the pinned EM (in contrast to five visible humps in the CPDs of a 
rotating electron molecule). Lengths in units of $l_B$. The units of the vertical axes are 
$10^{-2} l_B^{-2}$. The electron densities are normalized to the number of particles, $N$.
}
\label{n6nu13super}
\end{figure}

To demonstrate that weak pinning leads to formation of a nonrotating Wigner-crystal-type state,
we display in Fig.\ \ref{n6nu13super} the electron densities for (a) $\Phi^{\text{PIN}}
(45, 50; 1/\sqrt{2},1/\sqrt{2})$ and (b) $\Phi^{\text{PIN}}(45, 51; 1/\sqrt{2},1/\sqrt{2})$, 
where EXD yrast states have been used for both $\Phi_{L_1}$ and $\Phi_{L_2}$. Fig.\ 
\ref{n6nu13super} demonstrates that the pinning of LLL states leads to formation of explicitly 
nonrotating EMs with the molecular configurations being present in the electron densities 
themselves. This amounts to a ``reverse projection'' \cite{note7} $-$ that is, construction of
a symmetry-broken, nonrotating, pinned state via superposition of symmetry-conserving,
liquid-like states (with good total angular momenta), which themselves are characterized by
azimuthally uniform electron densities [see Figs.\ \ref{n6nu13exd}(a-c)], but exhibit intrinsic 
crystalline correlations manifested in the corresponding CPDs [see Figs.\ \ref{n6nu13exd}(d-f)].

\begin{figure}[t]
\centering\includegraphics[width=7.4cm]{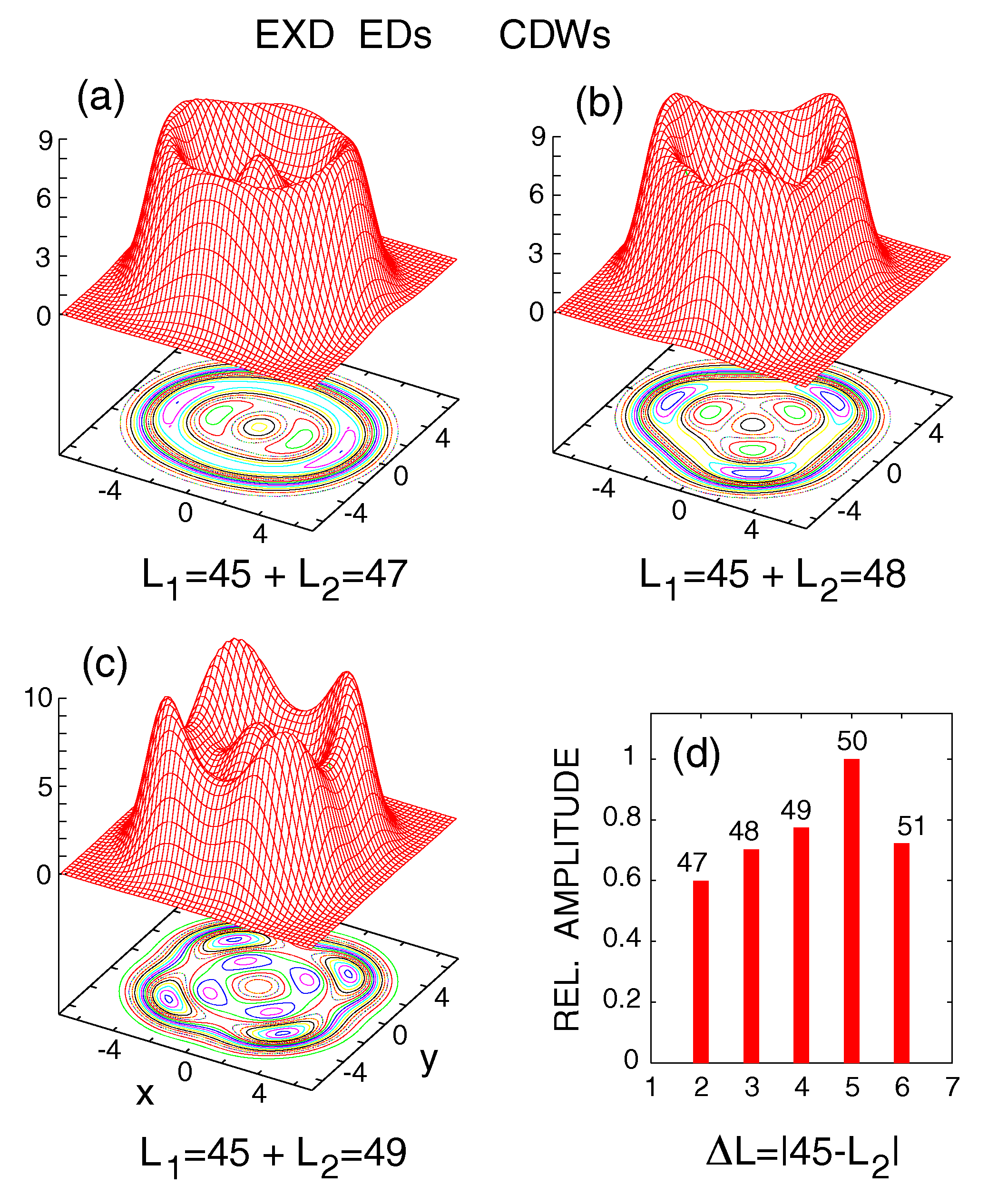}
\caption{(Color online) Electron densities corresponding to charge density waves for pinned 
[see Eq.\ (\ref{pinnwf})] LLL states in the neighborhood of $\nu=1/3$ for $N=6$ electrons
(corresponding to angular momenta around $L=45$). EXD states have been used for both 
$\Phi_{L_1}$ and $\Phi_{L_2}$. (a) $L=45$ and $L_1=47$ ($|L_1-L_2|=2$). (b) $L_1=45$ and 
$L_2=48$ ($|L_1-L_2|=3$). (c) $L_1=45$ and $L_2=49$ ($|L_1-L_2|=4$). Unlike the cases in Fig.\ 
\ref{n6nu13super}, the EDs here are not commensurate with the (1,5) or (0,6) classical 
Wigner-molecule equilibrium configurations; instead they represent charge density waves. 
(d) The relative amplitude of the density oscillations as a 
function of $\Delta L= |45-L_2|$, referenced to the WM case with $\Delta L=5$ [which is the 
strongest one, see Fig.\ \ref{n6nu13super}(a)]. The numbers above the vertical bars denote the 
values of $L_2$. Lengths in units of $l_B$. The units of the vertical axes are 
$10^{-2} l_B^{-2}$. The electron densities are normalized to the number of particles, $N$.
}
\label{n6superexci}
\end{figure}

A remarkable trend revealed by the EDs in Fig.\ \ref{n6nu13super} is that the (1,5) molecular 
configuration corresponds to a superposition of two EXD yrast states with angular momenta 
differing by $|L_1-L_2|=5$ angular momentum units, while the (0,6) molecular configuration 
corresponds to a superposition of two EXD yrast states with angular momenta differing by 
$|L_1-L_2|=6$ units. This motivated us to study the ED patterns in the neighborhood of $\nu=1/3$
for the superposition of two EXD yrast states as a function of the difference $\Delta L=
|L_1-L_2|$ (in particular for $\Delta L =2,3$ and 4; see Fig.\ \ref{n6superexci}). Of course 
the cases portrayed in Fig.\ \ref{n6superexci} involve mixing with excited states, which are
separated from the global ground state ($L_1=45$) by larger energy gaps (see Fig.\ 
\ref{n6glbspec}) and are not expected to materialize in a weak-pinning situation. 
Fig.\ \ref{n6superexci} illustrates that these combinations lead to formation of charge density
waves (CDWs) instead of Wigner-molecular crystallites (as in Fig.\ \ref{n6nu13super}). The
relative amplitudes of the oscillations in the EDs shown in Figs.\ \ref{n6superexci}(a-c), 
referenced to the WM with $\Delta L=5$ [Fig.\ \ref{n6nu13super}(a)], are shown in Fig.\
\ref{n6superexci}(d) exhibiting attenuated variations in the density.

\begin{figure}[t]
\centering\includegraphics[width=7.4cm]{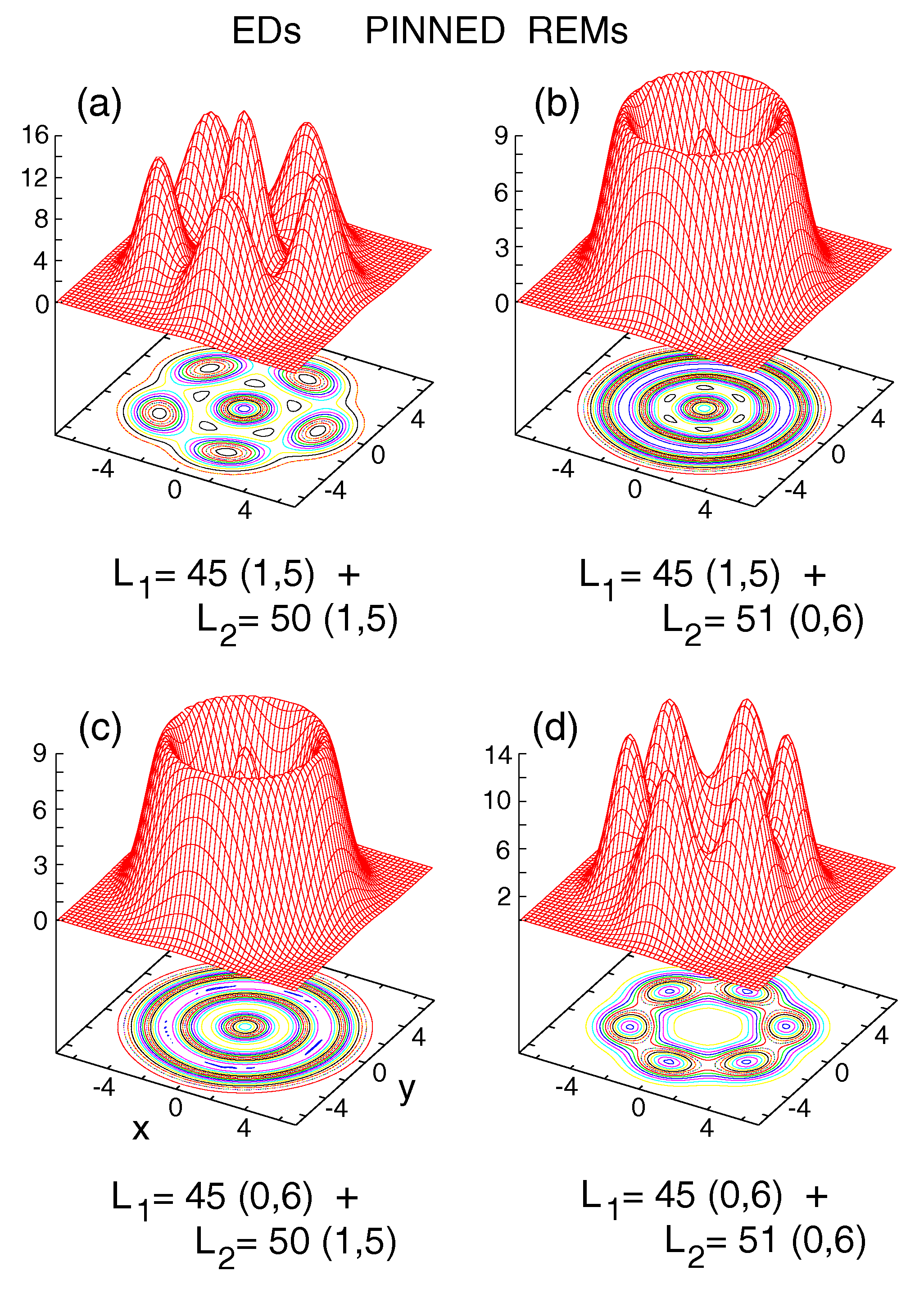}
\caption{(Color online) Electron densities for pinned [see Eq.\ (\ref{pinnwf})] LLL states in
the neighborhood of $\nu=1/3$ for $N=6$ electrons. Pure REM states 
$\Phi^{\text{REM}}_L(n_1,n_2)$  have been used for forming the superposition in Eq.\ 
(\ref{pinnwf}). (a) $L_1=45$ and $L_2=50$; with a (1,5) molecular isomer. (b) $L_1=45$ 
(1,5) and $L_2=51$ (0,6). (c) $L_1=45$ (0,6) and $L_2=50$ (1,5). (d) $L_1=45$ and $L_2=51$; 
both with a (0,6) isomer. The formation of a pinned (static) Wigner molecule representing a 
(1,5) molecular configuration in (a) and a (0,6) molecular configuration in (d) is 
apparent. Superpositions of REM functions belonging to different isomers [(b) and (c)] fail 
to produce a pinned crystalline structure. $\alpha=\beta=1/\sqrt{2}$. The units of the vertical
axes are $10^{-2} l_B^{-2}$. The electron densities are normalized to the number of particles, 
$N$.
}
\label{n6nu13superrvm}
\end{figure}

Fig.\ \ref{n6nu13superrvm} portrays the electron densities associated with superposition
of pure REM wave functions $\Phi^{\text{REM}}_L(1,5)$ and $\Phi^{\text{REM}}_L(0,6)$ for 
$N=6$ electrons. Such REM functions are the strongest components in the RVEM expansions of the
EXD LLL states for $L=45$, $L=50$, and $L=51$. We note that the magic angular momentum $L=45$
is commensurate with both the (1,5) and (0,6) isomeric structures, while the magic $L=50$ and
$L=51$ are commensurate only with one isomer, i.e., the (1,5) for the former and the (0,6) 
for the latter. Fig.\ \ref{n6nu13superrvm} shows that superposition of same-configuration RVEM 
functions leads to pinned WMs [see Figs.\ \ref{n6nu13superrvm}(a) for (1,5) and 
\ref{n6nu13superrvm}(d) for (0,6)]. In contrast, superposition of RVEM functions corresponding 
to different isomers fails to generate any crystalline structures. This means that the 
emergence of the EXD pinned crystallites (such as in Fig.\ \ref{n6nu13super}) cannot be 
explained (or anticipated) without the prior knowledge of the presence of appropriate 
$(n_1,n_2,...,n_r)$ isomeric RVEM functions as physical components in the EXD LLL states 
with good $L$ [see, e.g, the RVEM expansions in TABLE I and TABLE II, and in Ref.\ 
\onlinecite{yl10}].

The above considerations culminate in the following ``selection rules'' for the construction
of pinned Wigner-molecule crystallites: (I) The difference between the angular momenta of the
superimposed states [e.g., $L_1$ and $L_2$ in Eq.\ (\ref{pinnwf})] should be a multiple
of the magic angular momentum period associated with cusp states [e.g., either $\Delta L=5$ or 
6 for $N=6$, see Fig.\ \ref{n6LLLspec}(a)]. (II) A given isomer [e.g., (1,5) or (0,6) for
$N=6$] must have a participation weight in both the symmetry-conserving superimposed states
(for the participation weights for $N=6$ and $L=45$, see TABLE \ref{rvmdiagn6l45}). Naturally,
different participation weights give rise to different strengths of the density oscillations
in the Wigner-molecule crystallites [see Fig.\ \ref{n6superexci}(d)]. 

\begin{figure}[t]
\centering\includegraphics[width=8.0cm]{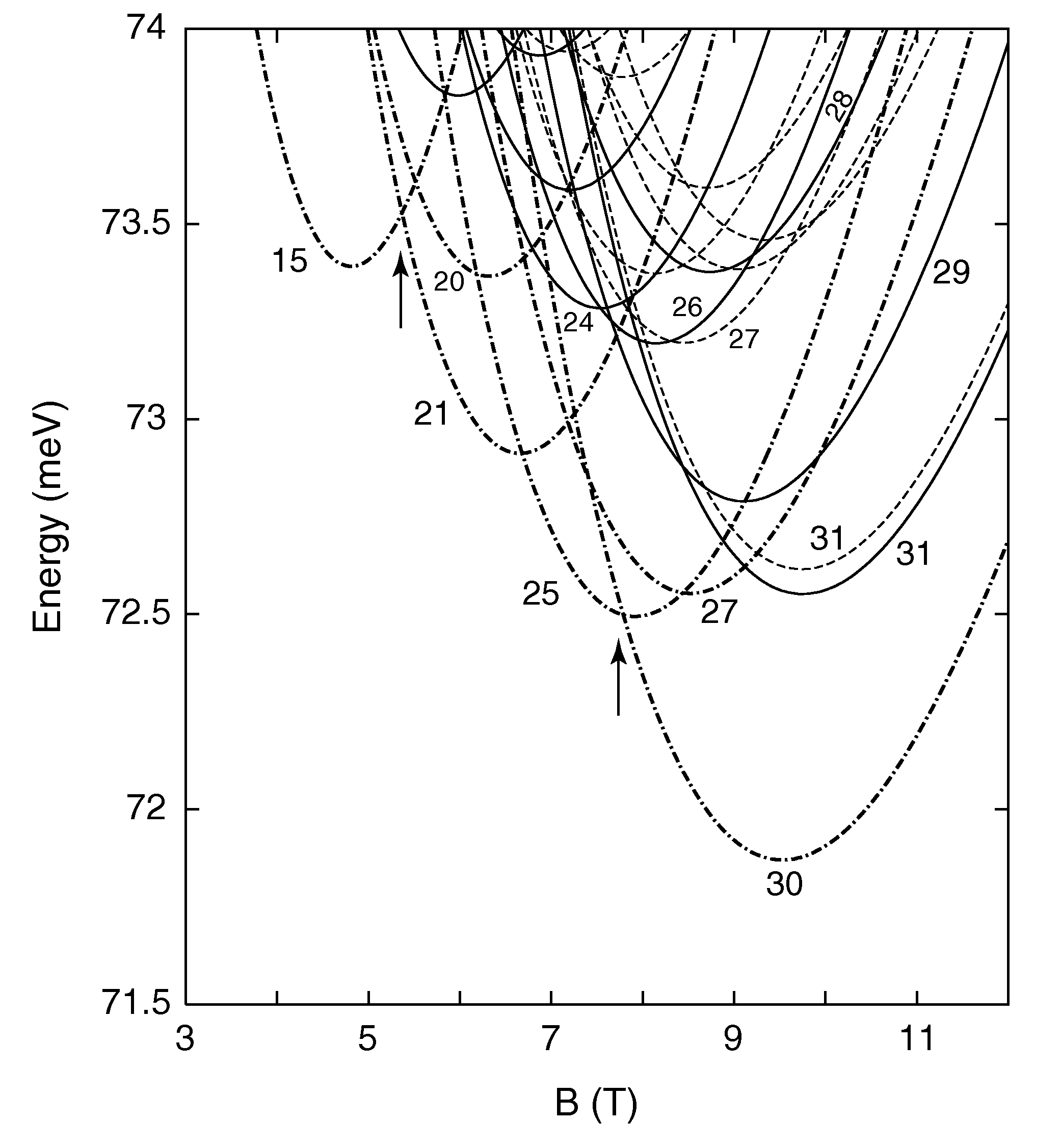}
\caption{The global spectrum as a function of the magnetic field in the
neighborhood of $\nu=1$ for $N=6$ LLL electrons. The confinement was taken as
$\hbar \omega_0 = 3.6$ meV. Note that all global ground states are yrast cusp 
states, but not all cusp states become global ground states.
\cite{jain95,yl03,lyl06} The yrast cusp states with $L=15$, 20 (1,5), 21 (0,6), 25 (1,5), 
27 (0,6), 30 (1,5) are portrayed by thick dashed-dotted lines. The $L=15$ curve relates to
$\nu=1$ in the thermodynamic limit. The numbers next to some curves denote the corresponding 
total angular momenta. Yrast states in addition to the cusp states are portrayed by a solid
line. The arrows highlight a couple of curve crossings (for curves associated with cusp states).
Remaining parameters: $\kappa=13.1$ and $m^*=0.067 m_e$, corresponding to GaAs. The topology 
(relative position) of the curves is independent of the specific value for $\hbar \omega_0$ 
(see text).
}
\label{n6glbspec15}
\end{figure}

\section{Pinned electron molecule in the neighborhood of $\nu=1$}
\label{secpinnu1}

The recent experimental observation in the neighborhood of $\nu=1$ (in addition to the 
$\nu=1/3$ neighborhood) of a microwave resonance in the spectrum of a 2D electon system under
high $B$ has also been associated with the formation of a weakly-pinned Wigner solid. 
\cite{zhu10.2,zhu10.1} In this Section, following an analysis similar to that used in Sec.\ 
\ref{secpinn} for the $\nu=1/3$ neighborhood, we show that crystallite states (precursors to
a Wigner solid in the thermodynamic limit) develop also naturally for a finite system in the 
neighborhood of $\nu=1$.

We start by displaying in Fig.\ \ref{n6glbspec15} the global spectrum for $N=6$ electrons; in 
this case $L=L_0=15$ corresponds to an integral filling factor $\nu=1$ [see Eq.\ (\ref{nu})].
The global LLL spectrum around $\nu=1$ (Fig.\ \ref{n6glbspec15}) shares the same prominent 
characteristics with that in the neighborhood of $\nu=1/3$ (Fig.\ \ref{n6glbspec}), i.e., only 
yrast cusp states associated with magic angular momenta (here $L=15$, 21, 25, 30) can become 
global ground states for a given magnetic field. The rest of the LLL states [derived from the 
interaction-only Hamiltoninan $H^{\text{int}}_{\text{LLL}}$; see Eq.\ (\ref{hlll})] become 
excited states in Fig.\ \ref{n6glbspec15} and they are separated by substantial gaps from the 
global ground states.

\begin{figure}[t]
\centering\includegraphics[width=8.4cm]{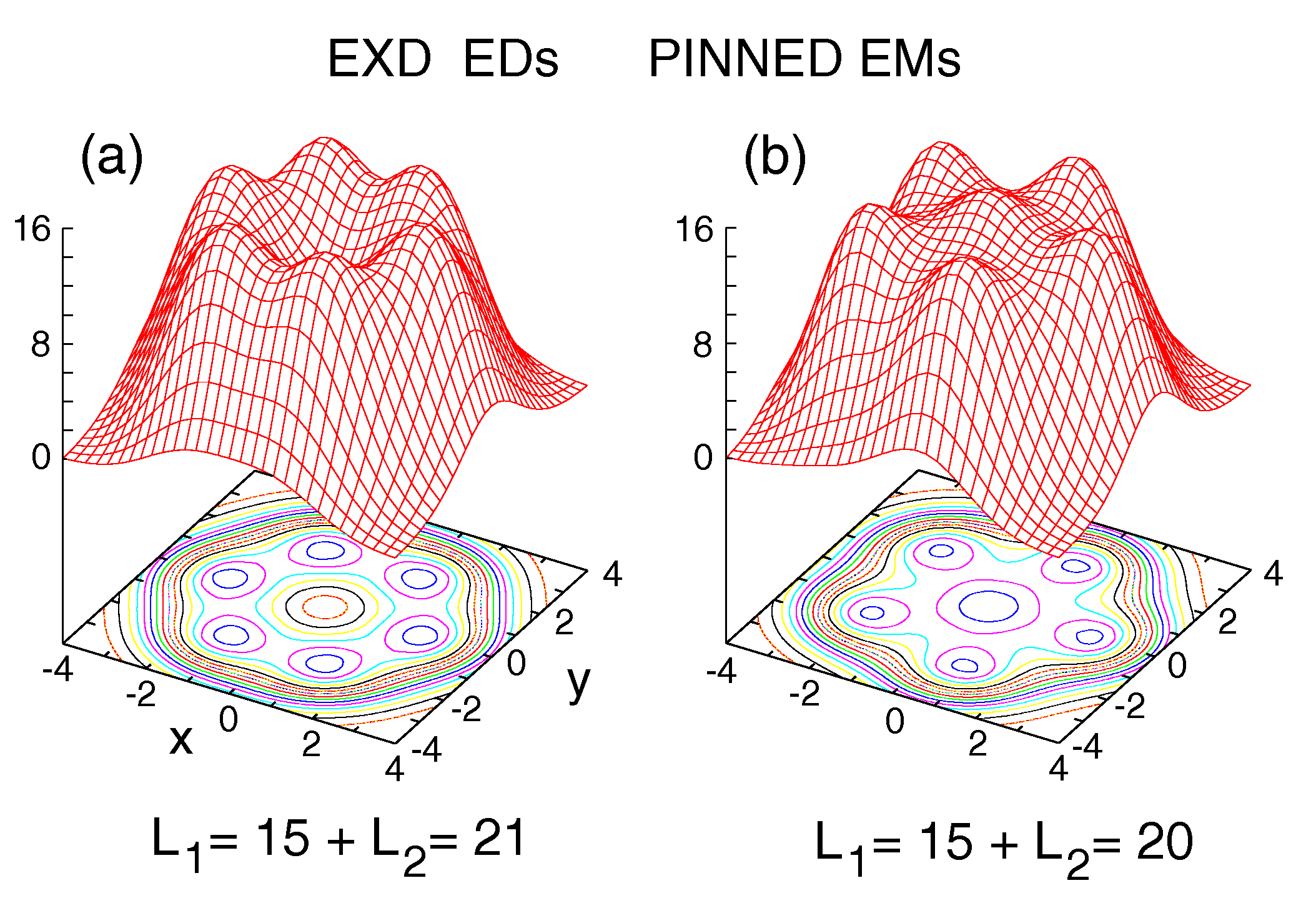}
\caption{(Color online) Electron densities for pinned [see Eq.\ (\ref{pinnwf})] LLL states in
the neighborhood of $\nu=1$ $(L=15)$ for $N=6$ electrons. EXD states have been used for both
$\Phi_{L_1}$ and $\Phi_{L_2}$. (a) $L_1=15$ and $L_2=21$ ($|L_1-L_2|=6$). (b) $L_1=15$ and 
$L_2=20$ ($|L_1-L_2|=5$). The formation of a pinned (nonrotating) EM representing a (0,6) 
molecular configuration in (a) and a (1,5) molecular configuration in (b) is transparent. 
$\alpha=\beta=1/\sqrt{2}$. Lengths in units of $l_B$. The units for the vertical axes are 
$10^{-2} l_B^{-2}$. The electron densities are normalized to the number of particles, $N$. 
}
\label{n6nu1super}
\end{figure}

As was pointed out in Sec.\ \ref{secpinn}, weak pinning results in the mixing of
two global-ground states in the neighborhood of crossing points (see, e.g., arrows in
Fig.\ \ref{n6glbspec15}), according to the prescription in Eq.\ (\ref{pinnwf}). 
To demonstrate that weak pinning leads to formation of a nonrotating Wigner-crystal-type state
in the neighborhood of $\nu=1$, we display in Fig.\ \ref{n6nu1super} the electron densities 
for (a) $\Phi^{\text{PIN}} (15, 21; 1/\sqrt{2},1/\sqrt{2})$ and (b) 
$\Phi^{\text{PIN}}(15, 20; 1/\sqrt{2},1/\sqrt{2})$, where EXD yrast states have been used for 
both $\Phi_{L_1}$ and $\Phi_{L_2}$. Fig.\ \ref{n6nu1super} shows that the pinning of LLL states
in the neighborhood of $\nu=1$ leads also to formation of explicitly nonrotating EMs, 
with the molecular configurations being present in the very electron densities.
We stress again the property that a difference of $\Delta L=6$ in angular momenta generates a 
(0,6) isomer, while a difference of $\Delta L=5$ generates a (1,5) isomer; this was also the 
case in the neighborhood of $\nu=1/3$ (see Fig.\ \ref{n6nu13super} and the selection rules
given at the end of Sec.\ \ref{secpinn}).

We further note that the filling factor corresponding to the crossing point of 
the $L_1=15$ and $L_2=21$ global curves is 0.857, while the filling factor for the crossing 
point of the $L_1=45$ and $L_2=50$ global curves is 0.316; here the filling factor is calculated 
from Eq.\ (\ref{nu}). As a result, we find that the pinned Wigner crystallite can appear in a 
larger range of filling factors away from $\nu=1$ (i.e., $\Delta \nu = 0.143$) compared to the 
corresponding range in the neighborhood of $\nu=1/3$ (where $\Delta \nu = 0.333 - 0.316 = 0.017$). 
This trend  is in agreement with experimental observations. \cite{zhu10.2}  

\section{Larger sizes}
\label{seclarg}

In the previous sections, we addressed the interplay of liquid and crystalline states by studying in
detail EXD results for the case of $N=6$ electrons. Our findings, however, are not limited to the 
case of $N=6$ electrons, but extend to larger sizes; this is supported by the EXD results presented 
in this section for sizes in the range from $N=7$ to $N=29$ electrons. 

\subsection{Extrapolation of total energies}
\begin{figure}[t]
\centering\includegraphics[width=8.4cm]{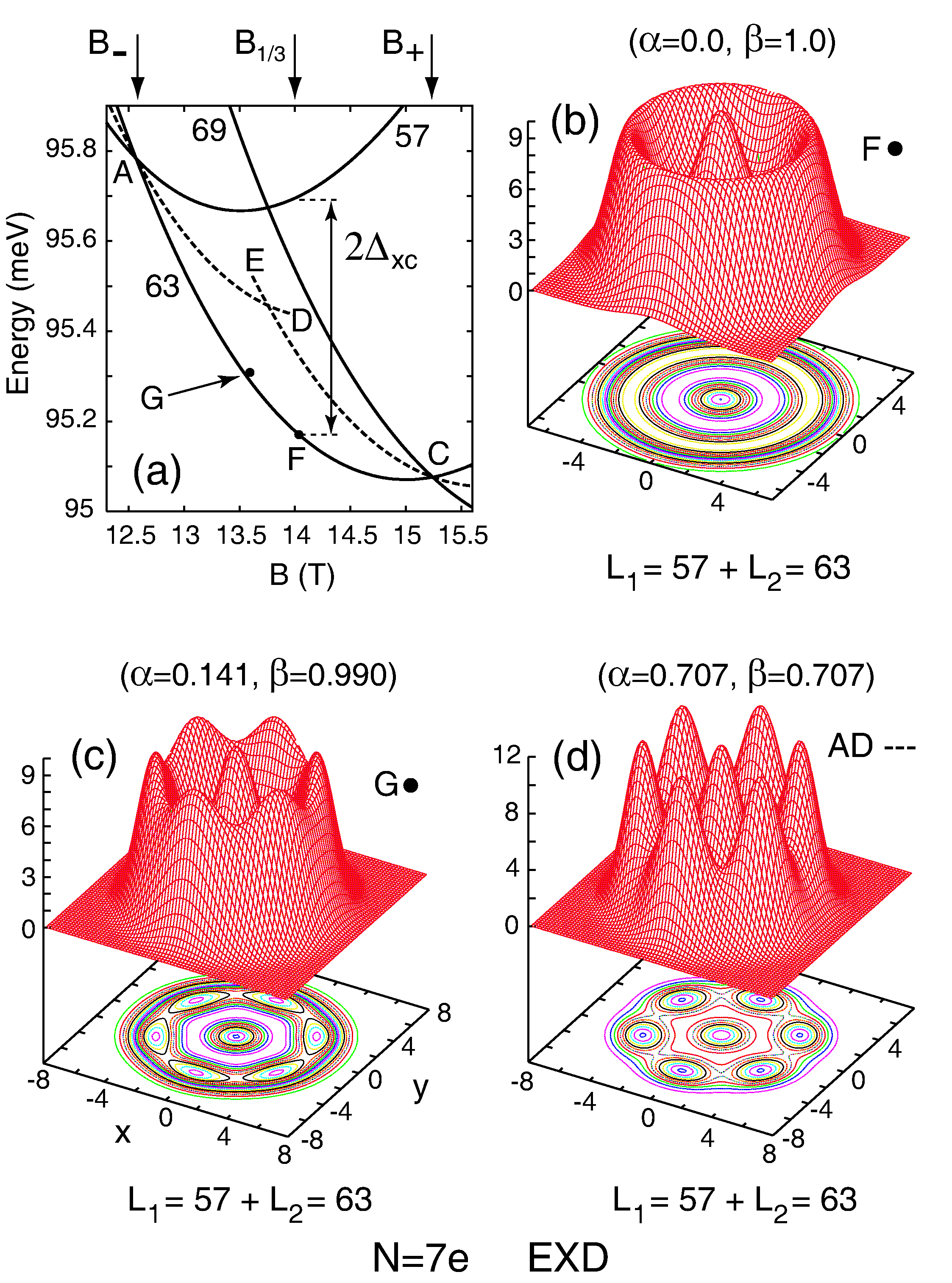}
\caption{(Color online) (a) The global energy spectrum [ground state ($L=63$) and first excited states
($L=57$ and $L=69$)] at $\nu=1/3$ (and its neighborhood) for $N=7$ electrons, as a function of the 
magnetic field $B$. The magnetic field corresponding to filling factor $\nu=1/3$ is denoted by $B_{1/3}$,
and those corresponding to the crossing points $A$ and $C$ are denoted by $B_-$ and $B_+$. The dashed 
lines indicate the energies for the pinned crystalline states [Eq.\ (\ref{pinnwf})] 
$\Phi^{\text{PIN}}(57, 63; 1/\sqrt{2},1/\sqrt{2})$ ($AD$, left dashed line) and 
$\Phi^{\text{PIN}}(63, 69; 1/\sqrt{2},1/\sqrt{2})$ ($EC$, right dashed line). (b-d) These panels 
portray the elecrton density of the pinned crystallite $\Phi^{\text{PIN}}(57, 63; \alpha,\beta)$ for 
various values of the weights $\alpha$ and $\beta$, corresponding to different degrees of pinning at 
the points marked in (a) as $F$ and $G$ and on the dashed-line segment marked in (a) as $AD$. In (a) 
the points marked $F$, $G$ and $A$ can be reached via a weak-pinning disorder, while moving from $A$ 
to $D$ along the dashed line would require strong-pinning disorder. The energy cost (gap) for mixing 
the ground state ($L=63$) with the first excited state ($L=57$) at $B_{1/3}$ [point marked $F$ in (a)]
is denoted as $\Delta_{\text{xc}}$. The confinement was taken as $\hbar \omega_0 = 3.6$ meV. Remaining
parameters: $\kappa=13.1$ and $m^*=0.067 m_e$, corresponding to GaAs. The topology (relative position)
of the curves is independent of the specific value for $\hbar \omega_0$ (see Sec.\ \ref{secpinn}). 
Lengths in units of $l_B$. The units of the vertical axes are $10^{-2} l_B^{-2}$. The electron density
is normalized to the number of particles, $N$.
}
\label{gapn7l63}
\end{figure}

In Fig.\ \ref{gapn7l63}(a), we plot the three lowest global ground-state energies around $\nu=1/3$ (for
$N=7$ electrons) as a function of the magnetic field $B$. They correspond to three cusp states 
with angular momenta 57, 63, and 69. The magnetic field corresponding
to $\nu=1/3$ is denoted by $B_{1/3}$, while those associated with the two crossing points $A$ and $C$
(left and right of $B_{1/3}$) are denoted as $B_-$ and $B_+$, respectively. The $AD$ dashed line
corresponds to the broken-symmetry (pinned) Wigner-crystallite state 
$\Phi^{\text{PIN}}(57, 63; 1/\sqrt{2},1/\sqrt{2})$ with energy $E^{\text{PIN}}=
E_{\text{glb}}(L_1=57)/2+E_{\text{glb}}(L_2=63)/2$, while the $EC$ dashed line corresponds to a pinned 
crystalline state $\Phi^{\text{PIN}}(63, 69; 1/\sqrt{2},1/\sqrt{2})$ with energy 
$E^{\text{PIN}}= E_{\text{glb}}(L_1=63)/2+E_{\text{glb}}(L_2=69)/2$. 

The energy cost (energy gap to be overcome) for mixing the $L=63$ ground state at $\nu=1/3$ 
with the excited state $L=57$ directly (vertically) above it, yielding the pinned crystallite 
$\Phi^{\text{PIN}}(57, 63; 1/\sqrt{2},1/\sqrt{2})$ is denoted by $\Delta_{\text{xc}}$ [see Fig.\ 
\ref{gapn7l63}(a)]. Most importantly, at $B_-$ (or $B_+$) the good-angular momentum states $L_1=57$ 
and $L_2=63$ (or $L_1=63$ and $L_2=69$) are degenerate and thus the energy cost (gap) for creating the 
crystallite from a superposition of two angular momenta states vanishes. The least favorable place 
(that is, the largest energy cost) for creating the crystallite is at $B_{1/3}$, while,
as aforementioned, at $B_-(B_+)$ the cost vanishes. Since
$\nu_{1/3} B_{1/3} = \nu_- B_- = \nu_+ B_+$ (keeping the electron density constant), this correlates 
with the experimentally observed continuous reduction of the microwave absorption strength as the filling
factor $\nu$ approaches the value 1/3, reflecting the enhanced stability of the liquid state
at $\nu=1/3$ compared to the crystalline one. As one moves away from $B_{1/3}$, the energy
cost for creating the crystallite decreases, so that a weaker disorder can act as a pinning
perturbation leading to the formation of the crystallite, as illustrated in Fig.\ \ref{gapn7l63}. 

\begin{figure}[t]
\centering\includegraphics[width=6.4cm]{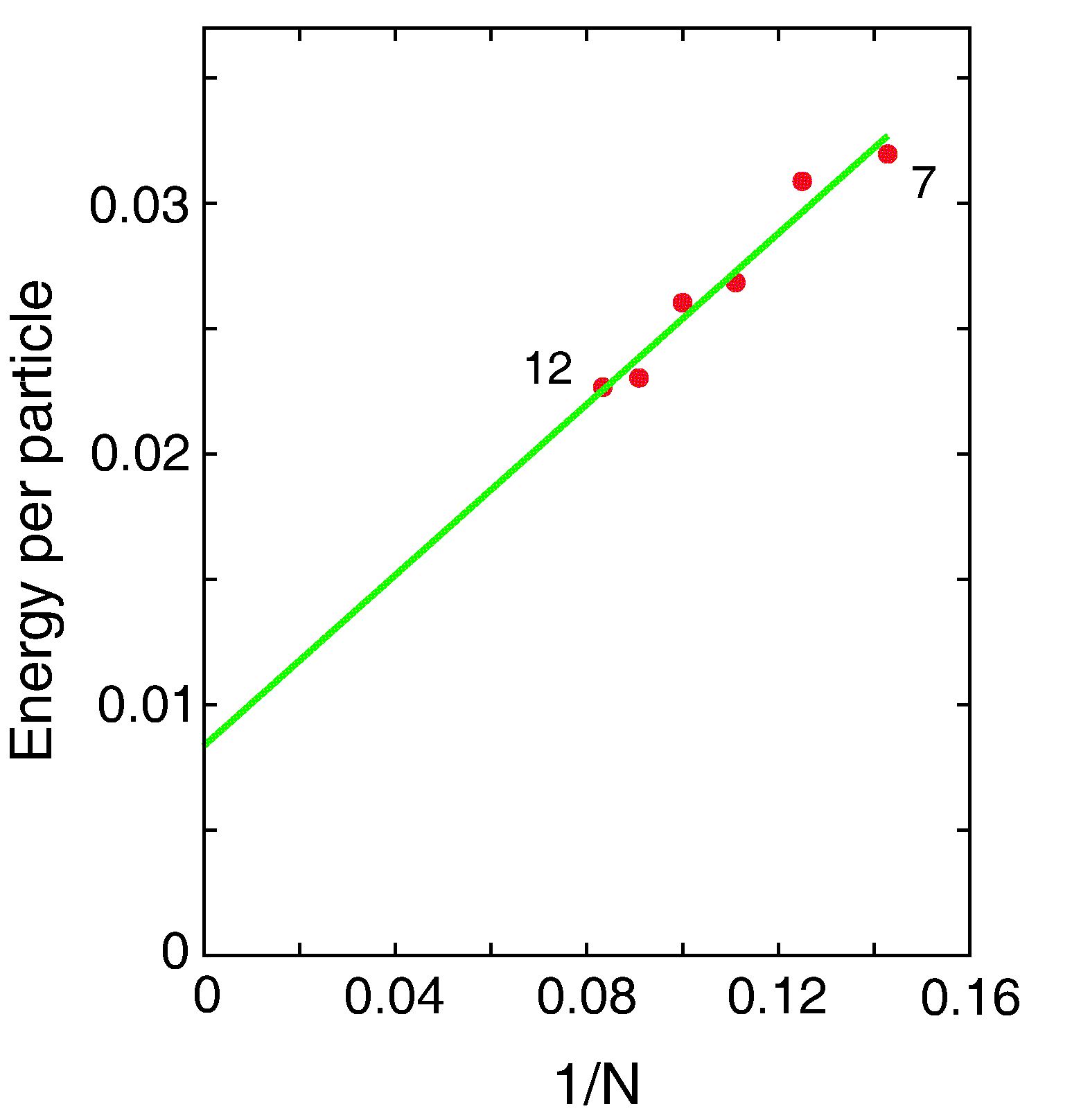}
\caption{(Color online) Extrapolation of the EXD-calculated [see Eq.\ (\ref{hlll})] energy gap per 
particle, $2 \Delta_{\text{xc}}/N$, at $\nu=1/3$ to the thermodynamic limit ($1/N \rightarrow 0$). 
Results are shown in the range $N=7$ to $N=12$ (see TABLE \ref{gapextra}). In the thermodynamic limit,
the energy cost per particle, $\Delta_{\text{xc}}/N$, to create a pinned crystal at precisely 
$\nu=1/3$ is approximately $0.004 e^2/\kappa l_B$. Vertical axis: Energies per particle in units of 
$e^2/\kappa l_B$. Horizontal axis: $1/N$ where $N$ is the number of electrons. 
}
\label{extra}
\end{figure}

The gradual development of a Wigner crystallite in the neighborhood of 1/3 for weak-pinning conditions
is illustrated in Fig.\ \ref{gapn7l63}(b-d), where the pinned state is illustrated at the points 
marked $F$, $G$, and $A$, respectively. To simulate the experimental finding of a liquid state at
$\nu=1/3$ $(B_{1/3})$, we assume a sufficiently weak pinning so that the weights $\alpha$ and $\beta$ in
the superposition $\alpha \Phi_{L_1=57} + \beta \Phi_{L_2=63}$ [see Eq.\ (\ref{pinnwf})] 
can be taken as $\alpha=0$ and $\beta=1$; indeed the electon density in Fig.\ 
\ref{gapn7l63}(b) is circularly symmetric corresponding to a liquid state. We remark that because of
the large mixing energy gap at $B_{1/3}$, creation of a pinned crystalline state at $\nu=1/3$ requires
strong-pinning disorder. The electron density associated with a pinned crystallite shown in Fig.\
\ref{gapn7l63}(c) [corresponding to the point marked $G$ in Fig.\ \ref{gapn7l63}(a)] was obtained via
weak-pinning induced mixing ($\alpha=0.141, \beta=0.990$). This electron density exhibits partially
developed crystalline features, with a (1,6) electronic configuration. A fully developed (1,6)
crystallite (obtained for $\alpha=\beta=1/\sqrt{2}$) is shown in Fig.\ \ref{gapn7l63}(d) [corresponding to
the point marked as $A$ in Fig.\ \ref{gapn7l63}(a)], which as aforementioned is associated with a
vanishing mixing gap (i.e., most susceptible to pinning by weak disorder).

\begin{table}[b] 
\caption{\label{gapextra}%
Interaction energies per particle [see the Hamiltonian in Eq.\ (\ref{hlll})] from $N=7$ to $N=12$ of the
yrast states entering in the evaluation of the gap $2 \Delta_{\text{xc}}/N$. $(n_1,n_2)$ denotes the ring
configuration. Energies in units of $e^2/\kappa l_B$.
}
\begin{ruledtabular}
\begin{tabular}{rrrrrr}
$N$ & $(n_1,n_2)$ & $L_1$ & $L_2$ $(3L_0)$ & $E^\text{int}_1/N$  & $E^\text{int}_2/N$    \\ \hline 
7   & (1,6)     &   57    &    63    &  0.57409   &  0.54213  \\
8   & (1,7)     &   77    &    84    &  0.63462   &  0.60373  \\
9   & (2,7)     &  101    &   108    &  0.68860   &  0.66177  \\
10  & (2,8)     &  127    &   135    &  0.74287   &  0.71684  \\
11  & (3,8)     &  157    &   165    &  0.79218   &  0.76915  \\                  
12  & (3,9)     &  189    &   198    &  0.84187   &  0.81921  \\                  
\end{tabular}
\end{ruledtabular}
\end{table}

To gain further insights into the nature of the Wigner crystalline states considered in this paper, it 
is instructive to extrapolate the EXD-calculated $\Delta_{\text{xc}}$ as a function of $1/N$ (where $N$ 
is the number of electrons) to the thermodynamic limit (i.e., $1/N \rightarrow 0$). Such extrapolation 
(see Fig.\ \ref{extra} and TABLE \ref{gapextra}) allows us 
to compare our results with previous treatments 
of the Wigner crystal based on variational wave functions in the bulk; \cite{maki83,lamg84,yi98} the 
latter results are summarized in TABLE \ref{encost}. Since the results from the bulk wave functions 
\cite{maki83,lamg84,yi98} assume that the kinetic energy of all the electrons is quenched to the value of
the LLL energy, $\hbar \omega_c$,  \cite{wexl03,jola10} we need to omit kinetic 
energy contibutions from $\Delta_{\text{xc}}$ when making comparisons; the energies used in Fig.\ 
\ref{extra} correspond to spectra like the one shown for $N=6$ in Fig.\ \ref{n6LLLspec}. Then 
$2\Delta_{\text{xc}}$ is given by the difference $|E^\text{int}_1-E^\text{int}_2|$ of the 
electron-electron interaction energies [the eigenenergies of the Hamiltonian in Eq.\ (\ref{hlll})] 
associated with the yrast state with (magic) angular momentum $L=3L_0=3N(N-1)/2$ ($\nu=1/3$) and the 
(magic) yrast state immediately preceeding it; see detailed description in TABLE \ref{gapextra}.  

\begin{table}[b] 
\caption{\label{encost}%
Energy cost per particle at the thermodynamic limit (compared to the liquid state) for forming a 
Wigner-crystalline state at $\nu=1/3$ according to previous approaches and the present work. Note
that smaller values reflect higher stability of the crystal. Values corresponding to previous 
Wigner-crystal approaches were extracted from Fig.\ 2 in Ref.\ \onlinecite{yi98}. Energies in units of 
$e^2/\kappa l_B$.
}
\begin{ruledtabular}
\begin{tabular}{lr}
Approach          & Energy cost per particle \\ \hline 
Maki-Zotos\footnote{Ref.\ \onlinecite{maki83}} (Hartree Fock)        &   0.0245            \\
Lam-Girvin\footnote{Ref.\ \onlinecite{lamg84}}                       &   0.0183            \\
Yi-Fertig\footnote{Ref.\ \onlinecite{yi98}} (composite-fermion WC)   &   0.0070            \\
Present Work\footnote{see Fig.\ \ref{extra}}                         &   0.0040            \\
\end{tabular}
\end{ruledtabular}
\end{table}

Inspection of the values in TABLE \ref{encost} leads us to conclude that the Wigner crystalline state 
described by our treatment entails the smaller gap (energy cost) of the crystal relative to the liquid 
state at $\nu=1/3$, compared to previous treatments. This finding is a consequence of the 
{\it quantum\/} nature of our crystalline state, exhibiting a high degree of electronic correlations. 
Since $\Delta_{\text{xc}}$ is largest at $\nu=1/3$ [see Fig.\ \ref{gapn7l63}(a)], the above conclusion
extends to the crystalline states formed (via weak-disorder pinning) in the whole neighborhood of 
$\nu=1/3$.

\begin{figure}[t]
\centering\includegraphics[width=8.4cm]{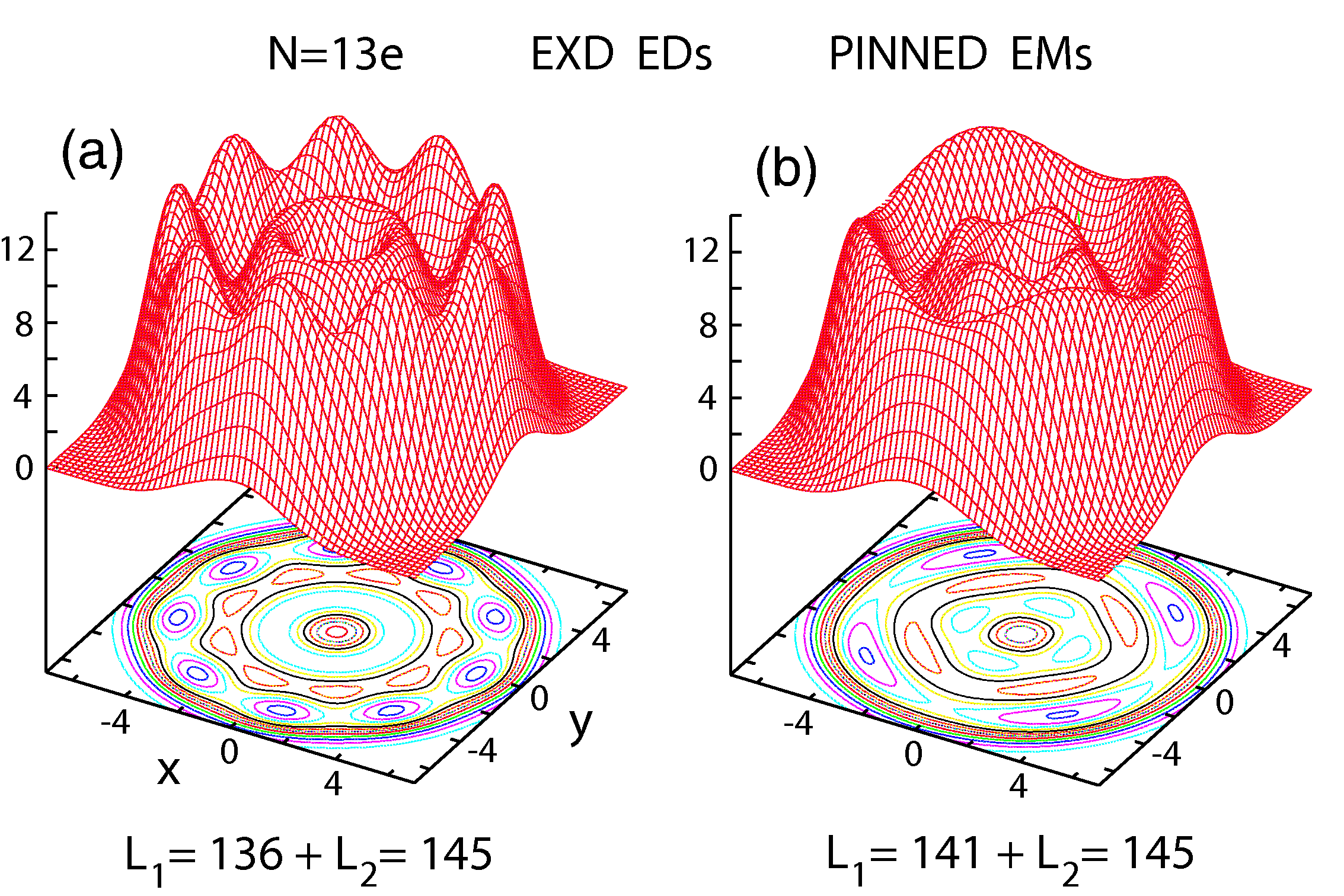}
\caption{(Color online) Electron densities for pinned (crystalline) [see Eq.\ (\ref{pinnwf})] LLL states 
in the neighborhood of $\nu=1$ for $N=13$ electrons. (a,b) Formation of a
pinned (nonrotating) EM representing a (4,9) molecular configuration is evident.
In (a) $L_2-L_1=9$, with the outer ring showing nine density humps, and in (b) $L_2-L_1=4$, showing four
density peaks on the inner ring.
$\alpha=\beta=1/\sqrt{2}$. Lengths in units of $l_B$. The units of the vertical axes are 
$10^{-2} l_B^{-2}$. The electron density is normalized to the number of particles, $N$.
}
\label{n13super}
\end{figure}

\begin{figure}[t]
\centering\includegraphics[width=8.4cm]{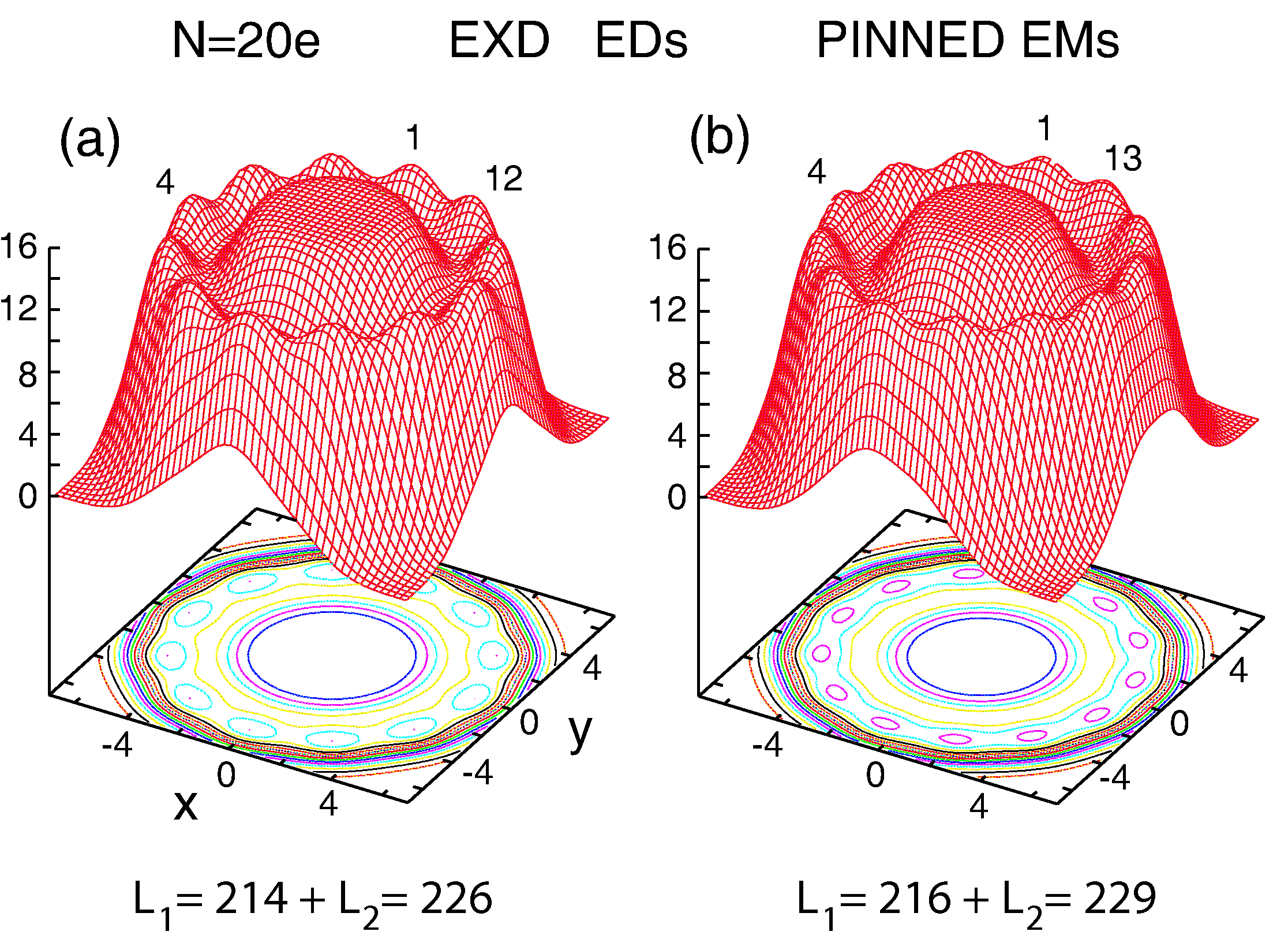}
\caption{(Color online) Electron densities for pinned [see Eq.\ (\ref{pinnwf})] LLL states in
the neighborhood of $\nu=1$ for $N=20$ electrons. In (a) $L_2-L_1=12$ corresponding to twelve electrons
on the outer ring, and in (b) $L_2-L_1=13$, with thirteen electrons on the outer ring. These
electron densities correspond to formation of pinned (nonrotating) EM isomers representing a 
(1,7,12) molecular configuration (a) and a (1,6,13) molecular configuration (b). 
$\alpha=\beta=1/\sqrt{2}$. Lengths in units of $l_B$. The units of the vertical axes are 
$10^{-2} l_B^{-2}$. The electron density is normalized to the number of particles, $N$.
}
\label{n20super}
\end{figure}

\subsection{Evolution of crystalline patterns}

In this section, we discuss the evolution of the pinned EXD crystalline patterns as a function of size 
(the number of electrons $N$). In Fig.\ \ref{gapn7l63}, in addition to the $N=6$ system discussed in
detail in earlier sections, we presented results for pinned Wigner crystallites in the neighborhood of 
$\nu=1/3$ for $N=7$ electrons; they conform to a (1,6) molecular configuration in agreement with
the finite-size crystalline structures for repelling classical point charges.\cite{beda94,kong03}

Currently, for $N > 10$, it is not computationally convenient to calculate electron densities (or CPDS)
in the neighborhood of $\nu=1/3$. However, given the fact that the crystalline isomeric structures are
independent of the filling factor $\nu$ (they depend only on the number of electrons $N$; compare Sec.\
\ref{secpinn} and Sec.\ \ref{secpinnu1}), we can use EXD results in the neighborhood of $\nu=1$ to study 
the evolution of pinned crystallites with size, without loss of generality.
   
\begin{figure}[t]
\centering\includegraphics[width=8.4cm]{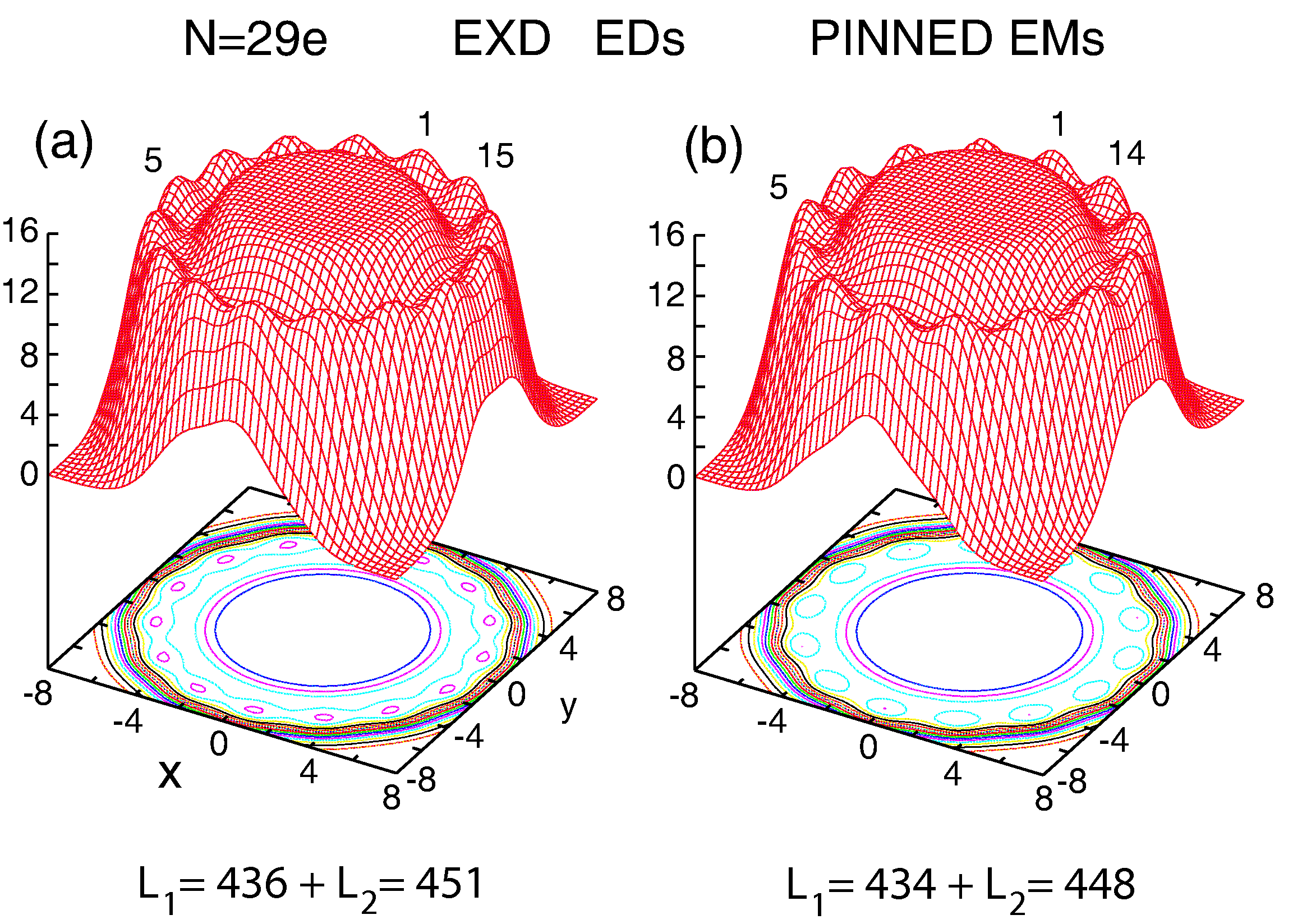}
\caption{(Color online) Electron densities for pinned (crystalline) [see Eq.\ (\ref{pinnwf})] LLL states 
in the neighborhood of $\nu=1$ for $N=29$ electrons. In (a) $L_2-L_1=15$ corresponding to fifteen
electrons on the outer ring, and in (b) $L_2-L_1=14$, with fourteen electrons on the outer ring. These
electron densities correspond to formation of pinned (nonrotating) EM isomers representing a
(4,10,15) molecular configuration (a) and a (5,10,14) molecular configuration (b).
$\alpha=\beta=1/\sqrt{2}$. Lengths in units of $l_B$. The units of the vertical axes are 
$10^{-2} l_B^{-2}$. The electron density is normalized to the number of particles, $N$.
}
\label{n29super}
\end{figure}

To this end, we present EXD calculated electron densities of pinned crystallites for three (larger than
$N=6$) sizes, i.e., $N=13$ (Fig.\ \ref{n13super}), $N=20$ (Fig.\ \ref{n20super}), and $N=29$ (Fig.\ 
\ref{n29super}).

In accordance with the selection rules described in Sec.\ \ref{secpinn}, the pinned EM for $N=13$ 
resulting from mixing states with $L_1=136$ and $L_2=145$ exhibits a 9-electron outer ring $(L_2-L_1=9)$
[Fig.\ \ref{n13super}(a)], and the one with $L_1=141$ and $L_2=145$ shows a 4-electron inner ring
$(L_2-L_1=4)$ [Fig.\ \ref{n13super}(b)]. The superposition of these two mixed states gives the (4,9) 
pinned configuration. For $N=20$ (Fig.\ \ref{n20super}) and $N=29$ (Fig.\ \ref{n29super}), we focus 
on the electron configurations in the outer rings of the pinned EM crystallites; the classical molecular
isomers\cite{beda94,kong03} exhibit the (1,7,12) and (1,6,13) crystalline configurations for $N=20$
and the (4,10,15) and (5,10,14) configurations for $N=29$. It is evident that the quantum mechanical
Wigner configurations of the outer rings in the EXD-calculated EDs in Fig.\ \ref{n20super} and 
Fig.\ \ref{n29super} are in agreement with the above classical patterns. For these sizes, i.e., $N=20$
and $N=29$, exploration of the molecular configurations of the electrons in the inner rings via EXD
calculations will require consideration of higher angular momenta and a heavier computational effort, 
beyond the scope of this paper.  

To summarize: For all sizes (in the range of $N=6$ to $N=29$) that we considered here (and for angular 
momentum values that we have been able to reach, at the present time, via quantum mechanical EXD 
calculations), the pinned crystalline configurations characterizing the electron densities are in 
agreement with those obtained from structural optimization of Coulomb repelling classical point charges 
confined by a 2D circular harmonic potential. \cite{beda94,kong03} This finding is particularly 
noteworthy since the LLL EXD wave functions are
determined solely by the interelectron repulsion [see the Hamiltonian in Eq. (\ref{hlll})].
As discussed in the context of the classical calculations (see TABLE I in Ref.\ \onlinecite{beda94}),
these configurations develop gradually a core that possesses a hexagonal Wigner-lattice structure for
larger clusters (above hundred particles). The aforementioned agreement supports the conjecture
that the quantum mechanical crystalline configurations described in this paper may be considered as 
embryonic Wigner crystallites extrapolating to the Wigner haxagonal lattice at the thermodynamic limit.  

\section{Discussion: Composite-fermion-crystal approaches versus the Wigner solid in
the neighborhood of $\nu=1/3$}
\label{seccfc}

The concept of a composite-fermion Wigner crystal was described \cite{yi98} through the use of 
the wave function approach, i.e., by attaching Jastrow vortices (factors) to the Maki-Zotos 
\cite{maki83} Hartree-Fock-crystal wave function. Of relevance for our purposes here is Fig.\ 2
in Ref.\ \onlinecite{yi98}, where the energies of the CFWC are compared to those of the 
Laughlin liquid states \cite{laug83} in the range $0.10 \le \nu \le 0.35$ (which includes the 
FQHE fillings 1/7, 1/5, and 1/3). From this figure, \cite{yi98} it is evident that the CFWC 
energy lies far above the Laughlin-liquid energy in the neighbohood of $\nu=1/3$. On the other 
hand the CFWC energy is competitive with the Laughlin energy in the neighborhood of $\nu=1/5$, 
and it becomes lower than the Laughlin energy in the neighborhood of $\nu=1/7$. The above 
trends suggest that the CFWC wave function is a legitimate candidate for the case of the Wigner
solid in the neighborhood of $\nu=1/5$, but {\it not\/} for the Wigner solid recently observed 
\cite{zhu10.2} in the neighborhood of $\nu=1/3$. This is consistent with most of the subsequent 
studies \cite{chan05,chan06} associated, or related, to CF crystals; indeed, we are unaware of 
any CF crystal study that addressed the neighborhood of $\nu=1/3$.

The similarity between the IQHE and the FQHE was used in Ref.\ \onlinecite{goer04} to study
whether the reentrant IQHE behavior \cite{goer04.2} may occur also for CFs in higher CF Landau 
levels. According to this analogy, residual interactions between CF quasiparticles (that is excitations
of the CF fractional quantum Hall effect liquid) may lead to formation
of CF-solid phases, or to second-generation CF liquids. Ref.\ \onlinecite{goer04} employed the 
same Hamiltonian composite-fermion approach as Narevich {\it al.\/} \cite{nare01} to model the CF solid 
and CF liquid phases around the electronic fractional fillings 4/11, 6/17, and 4/19, which are higher 
than 1/3. Such an approach (employing a two-component picture, i.e., CF liquid and its excitations),
which has been noted in Ref.\ \onlinecite{zhu10.2}, contrasts with our approach where a single class
of wave functions is used for both the liquid and Wigner-solid states.

Of relevance to our paper here is the fact that Ref.\ \onlinecite{goer04} did not produce new CF results 
in the neighborhood of $\nu=1/3$ with respect to the previous composite-fermion Wigner-crystal 
studies \cite{yi98} of Yi and Fertig. Furthermore the Hamiltonian CF approach employed in Ref.\
\onlinecite{goer04} appears not to describe the neighborhood of $\nu=1/3$, since it is a weak-coupling 
perturbative method applicable \cite{note5} only to cases ``when a higher CF LL level ($p \geq 
1$) is {\it partially\/} filled''; it fails when the composite-fermion filling factor 
$(\nu_{\text{CF}})$ is close to an integer value (corresponding to a closed CF shell). Note 
that for an electronic filling factor $\nu \sim 1/3$, one has $\nu_{\text{CF}} \sim 1$. 

The above approaches were explicitly based on a bulk 2D system. However, the liquid-like 
composite-fermion trial functions were formulated \cite{jain89,jainbook} in the context of a 
finite system. This offered several advantages, an important one being the ability to perform
quantitative comparisons \cite{jainbook} with exact results (e.g., for energies) and wave 
functions (e.g., pair correlations and overlaps). The composite-fermion crystal in Refs.\ 
\onlinecite{chan06,chan05} (henceforth referred to as CFC, to distinguish it from the 
aforementioned infinite CFWC) represents an attempt to formulate a CF crystal for a finite 
system. The important new element in the CFC approach is the use of the correlated 
rotating-electron-molecule \cite{yl02,yl04} wave function in the place of the 
{\it uncorrelated\/} single Slater determinant employed in the CFWC of Ref.\ \onlinecite{yi98}. 
This substitution is nontrivial, and (in the framework of the CFC theory) it leads
to restoration of the fundamental symmetries of the many-body Hamiltonian (rotational and 
translational) and to the introduction of additional energy-lowering correlations; a direct 
consequence is that the CFC wave function can be tested against exact diagonalization 
calculations, due to the fact that it has a good total angular momentum, $L_{\text{CFC}}$.

Because of its use of the REM (which is nonvanishing only for magic angular momenta,
$L_m$), the CFC is limited solely to the FQHE filings, and cannot provide descriptions
in the neighborhood of fractional fillings. Furthermore, a serious shortcoming of the 
CFC is its inability (by construction) to be extended to $\nu=1/3$. Indeed the angular 
momentum, $L_{\text{CFC}}$, of the CFC is 
given by \cite{chan05,chan06}
\begin{equation}
L_{\text{CFC}}=N(N-1)p + L_{\text{REM}},
\label{lcfc} 
\end{equation}
where the first term on the right-hand side is $2L_0p$, $p$ is a nonnegative integer, and 
$L_{\text{REM}}$ is the REM angular momentum. At $\nu=1/3$, one needs to have 
$L_{\text{CFC}}=3L_0$, with $L_0$ [see Eq.\ (\ref{l0})] being the lowest angular momentum. Then
the only possible value for $L_{\text{REM}}$ is $L_0$ $(p=1)$. The REM, however, at 
$L_{\text{REM}}=L_0$ coincides \cite{yl04} with the single Slater determinant of the maximun 
density droplet, and the usual attachment of two CF vortices to this determinant yields 
\cite{jain89} the liquid Laughlin wave function for $\nu=1/3$. 

From the above discussion, it follows that the emergence of a Wigner solid in the neighborhood 
of $\nu=1/3$ has been a challenging open problem in the composite-fermion literature up to 
date. Based on the insights gained in this paper and the equivalence \cite{yl10} between the 
composite-fermion and the RVEM theories, we show (see Appendix \ref{appcf}) that CF wave
functions can be used to describe formation of Wigner crystallites through the pinning process 
introduced in Sec.\ \ref{secpinn}.

\section{Summary}
\label{secsumm}

Based on the rotating-and-vibrating electron-molecule theory \cite{yl02,yl10} (RVEM), and in 
conjunction with exact-diagonalization results, we presented a unified microscopic theory for 
the interplay between liquid and Wigner-solid states in the neighborhood of $\nu=1/3$, which 
was recently observed \cite{zhu10.2} experimentally. In the RVEM theory, the description of both
liquid and Wigner-solid states is achieved within the framework of a single class of 
variational wave functions; see Eqs.\ (\ref{mol_trial_wf} and (\ref{qlam}) and Refs.\ 
\onlinecite{yl02,yl10}. 

In the RVEM method, liquid characteristics of the FQHE states are associated with conservation 
of the symmetries of the Hamiltonian, in particular the total angular momentum of the RVEM 
wave functions. For example, the electron densities of the RVEMs are 
circularly symmetric as expected for liquid states [this is also in accordance with EXD 
results for all states of the LLL spectra, as illustrated for $N=6$ electrons in the 
neighborhood of $\nu=1/3$ in Fig.\ \ref{n6nu13exd}(a-c)]. The liquid characteristics of the LLL 
states, however, coexist with intrinsic correlations that are crystalline in nature [i.e., 
exhibit patterns associated with the equilibrium configurations of $N$ classical point-like 
electrons, as revealed via the conditional probability distributions; see examples in Fig.\ 
\ref{n6nu13exd}(d-f)]. Further insight into the intrinsic crystalline correlations was gained 
via a study of the relative weights of the ro-vibrational excitations of the electron
crystallite. For $N=6$ electrons, examples of such relative weights were presented for states 
in the neighborhood of $\nu=1/3$ in Secs.\ \ref{secn6l45} and \ref{secn6l47}.

Although the electron densities of the symmetry-conserving LLL states do not exhibit 
crystalline patterns, the intrinsic crystalline correlations are reflected in the emergence in 
the LLL spectra of cusp yrast states with enhanced stability and magic angular momenta (see 
Fig.\ \ref{n6LLLspec}); the cusp states are associated with the fractional fillings in the 
thermodynamic limit [see Eq.\ (\ref{nu})]. A direct consequence of the enhanced stability is 
the fact that only states with magic angular momenta (cusp states) can become global ground 
states, as illustrated in Figs. \ref{n6glbspec} and \ref{n6glbspec15} for $N=6$ electrons and 
$\nu=1/3$ and $\nu=1$, respectively.

Away from the exact fractional fillings, weak pinning perturbations (experimentally due to weak
disorder) can overcome the energy gaps between adjacent global states (in particular near their
crossing points; see Fig.\ \ref{n6glbspec} and Sec.\ \ref{seclarg}) and generate a mixed, {\it 
broken symmetry\/} (pinned) ground state, that is a linear superposition of symmetry-conserving LLL 
states with different total angular momenta. A central finding of this paper is that such pinned 
states do exhibit explicitly a crystalline pattern in the electron density (nonrotating, pinned 
molecular, or Wigner, crystallites); see, e.g., Figs.\ \ref{n6nu13super} and \ref{n6nu13superrvm}). 
These pinned crystallites represent finite-size precursors of the Wigner solid in the thermodynamic 
limit (see Sec.\ \ref{seclarg}). Furthermore, we illustrated that the emergence of the pinned 
molecular crystallite is a direct consequence of the contributions of RVEM components in the 
symmetry-conserving LLL states themselves; see discussion in text related to Fig.\ 
\ref{n6nu13superrvm}. 

Along with the molecular crystallites (see Figs.\ \ref{n6nu13super} and \ref{n6nu13superrvm}),
other charge-density-wave patterns may develop, originating from the absence of certain
commensurability conditions between the angular momentum states that get coupled in the 
pinning process (see Fig.\ \ref{n6superexci}). However, they correspond to coupling of the
global ground states with excited global states, and therefore are less likely to materialize 
for a case of weak pinning, because of the large energy gaps between these states. Selection
rules governing the formation of pinned Wigner crystallites were formulated at the end of
Sec.\ \ref{secpinn}.

In addition to the neighborhood of $\nu=1/3$, we also demonstrated that the RVEM approach can 
account in a similar unified manner for the interplay between liquid and Wigner solid
states in the neighborhood of $\nu=1$; see Sec.\ \ref{secpinnu1}. 

We note again here that our findings are not limited to the case of $N=6$ electrons only.
In Sec.\ \ref{seclarg}, exact-diagonalization results were presented in a wide range 
of sizes, from $N=7$ to $N=29$ electrons. The extrapolation displayed in Fig.\ \ref{extra} gave
a value for the energy gap representing the stability of the bulk Wigner crystal. This value was
compared to previously calculated estimates by other methods in TABLE \ref{encost}; it was found 
to reflect a Wigner crystal of higher stability due to a large degree of quantum correlations.
Furthermore, we showed in Sec.\ \ref{seclarg} that the pinned crystalline patterns obtained via 
our quantum mechanical calculations evolve (for all sizes considered in this paper, i.e., from 
$N=6$ to $N=29$ electrons) according to the well established sequence of configurations found for 
classical point charges, leading to formation of Wigner-crystalline hexagonal cores for 
$N > 100$ electrons.

As aforementioned, the RVEM theory described here and applied to the analysis of the
appearance of Wigner crystalline patterns in the neighborhood of $\nu=1/3$ employs a single
class of variational wave functions for the description of both the correlated liquid 
and Wigner-solid states. This theory differs in an essential manner from composite-fermion 
approaches \cite{jainbook} (including Laughlin's original formulation \cite{laug83}) which 
utilize different classes of variational wave functions for representing the liquid versus 
Wigner-solid states. Specifically, in the CF approaches, FQHE states are associated with CF 
liquid states (defined in the context of $N$ LLL electrons and preserving the total angular 
momentum \cite{jain89,jainbook}), while solid states are described by CF Wigner crystals; 
\cite{yi98} the latter violate the conservation of the total angular momentum (broken symmetry)
and are a modification (the attachment of Jastrow factors) of the Maki-Zotos \cite{maki83} 
Wigner crystal for an infinite 2D system (defined on a triangular lattice at the mean-field 
Hartree-Fock level). \cite{note24}

In Sec.\ \ref{seccfc}, we discussed the open challenges faced by the composite-fermion
literature \cite{yi98,nare01,goer04,chan05,chan06,jainbook} in addressing the emergence of the
Wigner-solid state in the neighborhood of $\nu=1/3$.
Based on the insights gained in this paper and the equivalence \cite{yl10} between the
composite-fermion and the RVEM theories, we show (see Appendix \ref{appcf}) that CF wave
functions can be used to describe formation of Wigner crystallites through the pinning process
introduced in Sec.\ \ref{secpinn}.

The physical picture and formalism developed in this paper is expected to apply to other filling
fractions. While future experimental and theoretical studies will be needed, our work suggests that 
liquid-Wigner-solid coexistence may occur for fractions in the neighborhood of which a Wigner crystal 
has not been seen as yet. Investigations of these issues with a variable (tunable) degree of disorder
would be most valuable.

\begin{acknowledgments}
This work was supported by the Office of Basic Energy Sciences of the US D.O.E. 
under contract FG05-86ER45234.
\end{acknowledgments}

\begin{figure}[t]
\centering\includegraphics[width=8.4cm]{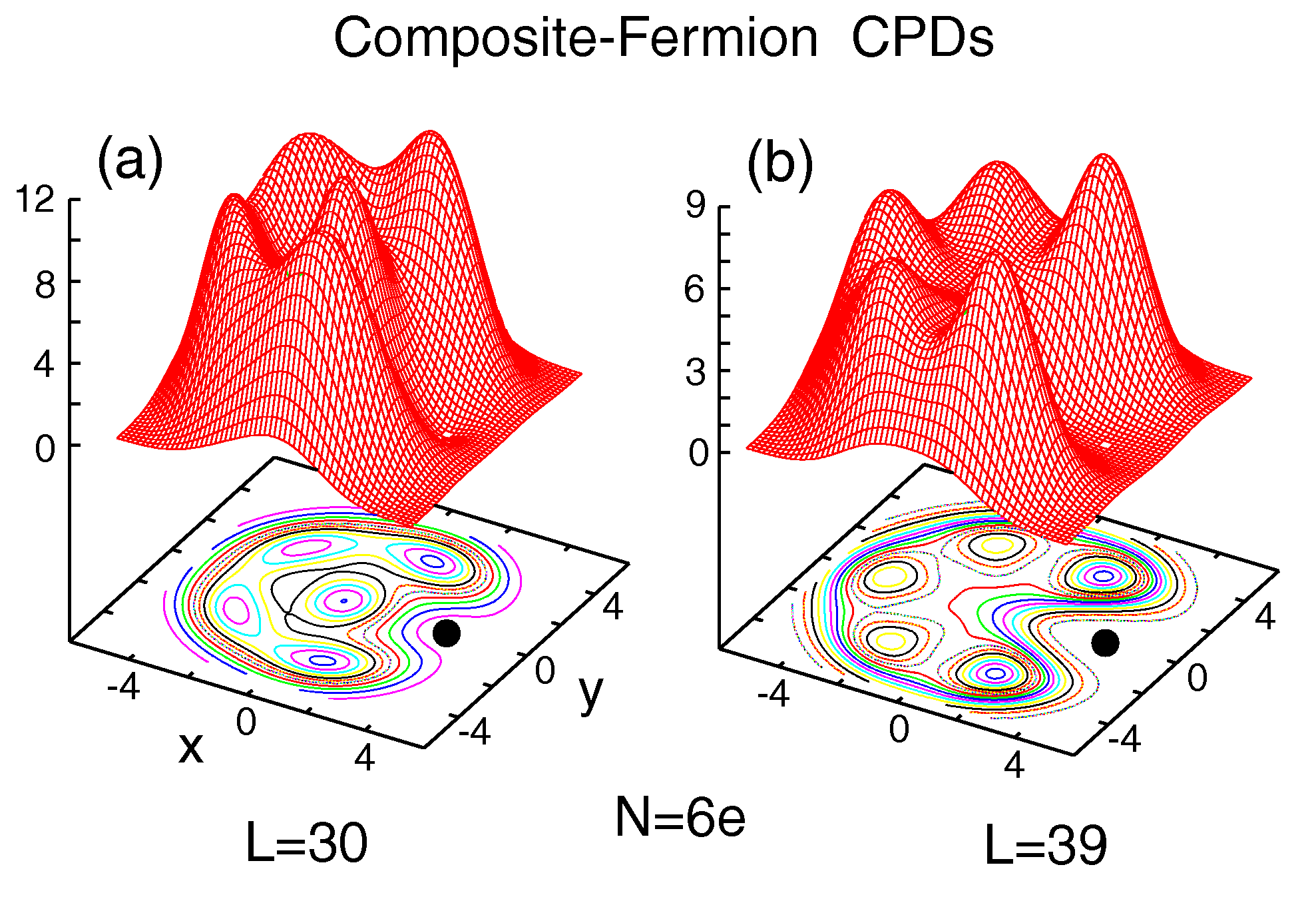}
\caption{(Color online) Composite-fermion CPDs for the cusp yrast states for $N=6$ LLL 
electrons with (a) $L=30$ ($\nu=1/2$) and (b) $L=39$. The compact CF trial 
functions for these $L$'s were calculated for a disk geometry according to Sec.\ 4.3 of Ref.\ 
\onlinecite{jain98}. The solid dots denote the position of the fixed point. The units for the 
vertical axes are arbitrary, but the same for all frames portraying CPDs throughout the paper. 
Lengths in units of $l_B$. Note the (1,5) and (0,6) molecular patterns for $L=30$ (a) and 
$L=39$ (b), respectively. 
}
\label{cpdn6cf}
\end{figure}

\appendix

\section{Purely rotational trial wave functions (REMs)}
\label{apprem}

In this Appendix, we recapitulate the analytic formulas for the vibrationless REM trial 
wave functions entering into the general expression for the RVEM functions [see Eq.\ 
(\ref{mol_trial_wf})]. The REM expresions for any $(n_1,n_2,\ldots,n_r)$ multi-ring 
configuration (with the number of electrons $N=\sum_{q=1}^r n_q$, $n_q$ being the number of
electrons in the $q$th ring) were derived earlier in Refs.\ \onlinecite{yl02,yl03,yl10}. 

Assuming that ${\cal L}_1$ and ${\cal L}_2$ are the {\it partial\/} angular momenta for
each ring (${\cal L}_1+{\cal L}_2={\cal L}$), the final two-ring $(n_1,n_2)$ REM expression is
\begin{eqnarray}
\Phi^{\text{REM}}_{\cal L} (n_1,n_2)[z] && \nonumber \\
&& \hspace{-2.3cm} 
= \sum_{0 \leq l_1 < l_2 < \ldots < l_{n_1} < l_{n_1+1} < \ldots < l_N}
^{l_1+l_2+\ldots+l_{n_1}={\cal L}_1, l_{n_1+1}+l_{n_1+2}+ \ldots +l_{N}={\cal L}_2}
C(l_1,l_2, \ldots, l_{n_1}) \nonumber \\
&& \hspace{-1.9cm} \times \; C(l_{n_1+1},l_{n_1+2}, \ldots, l_{N}) 
{\text{det}} [z_1^{l_1}, z_2^{l_2}, \ldots, z_N^{l_N}], 
\label{remn1n2}
\end{eqnarray}
where the $z_i$'s are complex-number particle coordinates and ``det'' denotes a Slater 
determinant. The coefficients $C(l_1,l_2, \ldots, l_{n_1})$ and 
$C(l_{n_1+1},l_{n_1+2}, \ldots, l_{N})$ are calculated by applying to each one of them 
the single-ring [$(0,N)$] expression
\begin{eqnarray}
C(l_1,l_2, \ldots, l_{N}) &=& \nonumber \\
&& \hspace{-3.2cm} \left( \prod_{i=1}^N l_i! \right)^{-1} 
\left( \prod_{1 \leq i < j \leq N} 
\sin \left[\frac{\pi}{N}(l_i-l_j)\right] \right).  
\label{coeff}
\end{eqnarray} 

It is straighforward to generalize the two-ring REM expression in Eq.\ (\ref{remn1n2}) 
to more complicated or simpler [i.e., $(0,N)$ and $(1,N-1)$] configurations by 
(I) considering a separate factor 
$ C(l_{n_{q-1}+1},l_{n_{q-1}+2}, \ldots, l_{ n_{q-1}+n_{q} })$ for each $q$th ring;
(II) restricting the summation of the associated $n_q$ angular momenta,
i.e., $l_{n_{q-1}+1}+l_{n_{q-1}+2}+ \ldots +l_{n_{q-1}+n_{q}}={\cal L}_q$, with
$\sum_{q=1}^r {\cal L}_q ={\cal L}$. 

\begin{figure}[t]
\centering\includegraphics[width=5.4cm]{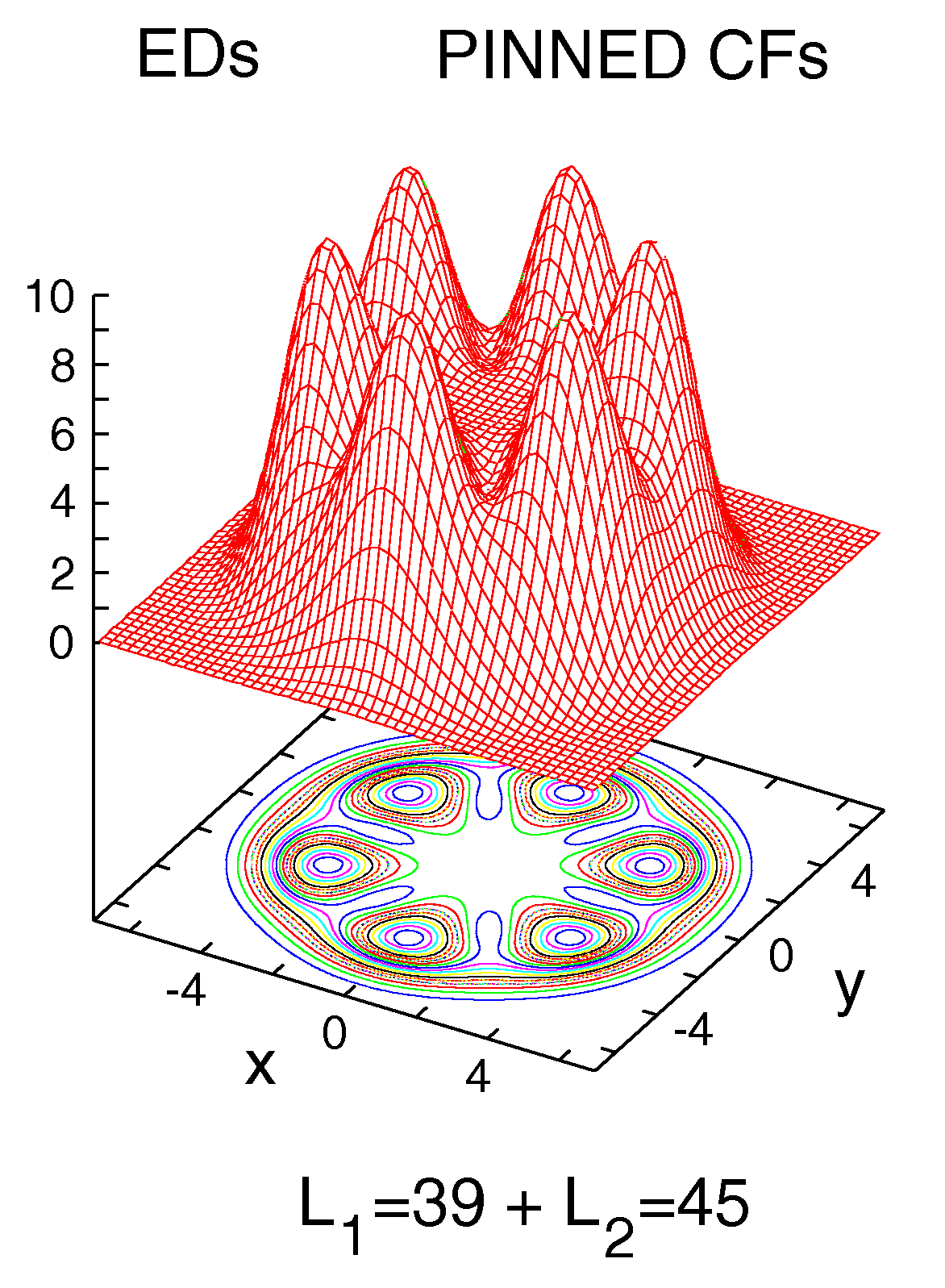}
\caption{(Color online) Electron densities for pinned [see Eq.\ (\ref{pinnwf})] LLL states in
the neighborhood of $\nu=1/3$ for $N=6$ electrons. Compact CF states have been used for both 
$\Phi_{L_1}$ and $\Phi_{L_2}$. $L_1=39$ and $L_2=45$ ($|L_1-L_2|=6$). The compact CF wave 
function at $L=45$ coincides with the Laughlin trial function. \cite{laug83} The formation of a
pinned (nonrotating) EM representing a (0,6) molecular configuration is transparent. 
$\alpha=\beta=1/\sqrt{2}$. Note that all six humps of localized electrons are visible in the
electron densities of the pinned CFs (in contrast to five visible humps in the CPDs in Fig.\
\ref{cpdn6cf}). Lengths in units of $l_B$. The units of the vertical axis are 
$10^{-2} l_B^{-2}$. The electron density is normalized to the number of particles, $N$.
}
\label{n6nu13supercf}
\end{figure}

The analytic expressions for $\Phi^{\text{REM}}_{\cal L}(n_1,n_2,\ldots,n_r)[z]$
describe pure molecular rotations associated with magic angular momenta
\begin{equation}
{\cal L}=L_m \equiv L_0+\sum_{q=1}^r n_q k_q, 
\label{mam}
\end{equation}
with $k_q$, $q=1,\ldots,r$ being nonnegative integers.

A central property of these trial functions is that identically
\begin{equation}
\Phi^{\text{REM}}_{\cal L}(n_1,n_2,\ldots,n_r)[z]=0
\label{eqzero}
\end{equation}
when 
\begin{equation}
{\cal L} \neq L_m
\label{neqtomagic}
\end{equation}
This selection rule follows directly from the point group symmetries of the 
$(n_1,n_2,\ldots,n_r)$ multi-ring polygonal configurations. Indeed under condition
(\ref{neqtomagic}) the $C(\ldots)$ coefficients are identically zero. In other words, purely 
rotational states are allowed only for certain angular momenta that do not conflict with the 
intrinsic molecular point-group symmetries.

\section{Intrinsic crystalline correlations in composite-fermion wave functions for $\nu > 1/5$}
\label{appcf}

Another class of trial functions that have been shown to approximate well (in energy) the EXD 
yrast cusp states are the composite-fermion ones; \cite{jainbook} here we refer in particular 
to the compact \cite{jain89} (also referred to \cite{jeon07} as mean-field) ones. For larger 
fractional fillings ($\nu > 1/5$, including $\nu=1/3$), it has been ascertained 
\cite{jainbook,jain00} that the compact CF functions represent paradigms of liquid states 
devoid of any intrinsic crystalline correlations. Since for $N=6$ electrons $\nu \geq 1/5$ corresponds
to angular momenta $L \leq 75$, the CF CPDs (for $L=30$ and $L=39$) displayed
in Fig.\ \ref{cpdn6cf}, however, disagree with the above assertion. (The CF wave functions were
calculated according to Sec.\ 4.3 of Ref.\ \onlinecite{jain98}.) Indeed, well formed 
crystalline correlations corresponding to the (1,5) molecular isomer (commensurate with a magic
angular momentum $L=30$) and the (0,6) molecular isomer (commensurate with a magic angular 
momentum $L=39$) are present in these CF CPDs. This is in agreement with the finding in Ref.\ 
\onlinecite{yl10} that all LLL functions with good $L$ are equivalent to rotating and vibrating
Wigner molecules.  

The above suggests that the superposition of CF wave functions should also yield pinned Wigner 
crystallites. This conclusion is explicitly confirmed in Fig.\ \ref{n6nu13supercf}, where the 
electron density (showing well developed crystalline oscillations) of a pinned CF state is 
displayed for a case in the neighborhood of $\nu=1/3$ [i.e., for a state constructed by mixing 
the compact CF states for $L_1=39$ and $L_2=45$, see Eq.\ (\ref{pinnwf})]. We note that the 
compact CF state for $N=6$ and $L=45$ coincides with the Laughlin trial function. \cite{laug83}
We further note that $L_1-L_2=6$, and that accordingly the crystalline configuration in Fig.\ 
\ref{n6nu13supercf} corresponds to the (0,6) classical molecular isomer. This demonstrates that
the selection rules for formation of Wigner crystallites (discussed at the end of Sec.\ 
\ref{secpinn}) apply to the CF trial functions as well.


\begin{thebibliography}{999}
\bibitem{tsui82}D. C. Tsui, H. L. Stormer, and A. C. Gossard,
Phys. Rev. Lett. {\bf 48}, 1559 (1982). 
\bibitem{laug83}
R. B. Laughlin,
Phys. Rev. Lett. {\bf 50}, 1395 (1983). 
\bibitem{fuku78}
H. Fukuyama and P. A. Lee, 
Phys. Rev. B {\bf 18}, 6245 (1978).
\bibitem{maki83}
K. Maki and X. Zotos,
Phys. Rev. B {\bf 28}, 4349 (1983).
\bibitem{tana89}
Liquid-to-solid (Wigner crystal) crossover as a function of electron density has been discussed,
using different variational wave functions for the two phases, for the two-dimensional electron gas
in the absence of a magnetic field; see B. Tanatar and and D. M. Ceperley, Phys. Rev. B {\bf 39},
5005 (1989).
\bibitem{lamg84}
P. K. Lam and S. M. Girvin, 
Phys. Rev. B {\bf 30}, 473 (1984).
\bibitem{andr88}
E. Y. Andrei, G. Deville, D. C. Glattli, F. I. B. Williams, E. Paris, and B. Etienne,
Phys. Rev. Lett. {\bf 60}, 2765 (1988).
\bibitem{li00}
C.-C. Li, J. Yoon, L. W. Engel, D. Shahar, D. C. Tsui, and M. Shayegan,
Phys. Rev. B {\bf 61}, 10905 (2000).
\bibitem{ye02}
P. D. Ye, L. W. Engel, D. C. Tsui, R. M. Lewis, L. N. Pfeiffer, and K. West, 
Phys. Rev. Lett. {\bf 89}, 176802 (2002).
\bibitem{chen04}
Y. P. Chen, R. M. Lewis, L. W. Engel, D. C. Tsui, P. D.
Ye, Z. H. Wang, L. N. Pfeiffer, and K. W. West, 
Phys. Rev. Lett. {\bf 93}, 206805 (2004).
\bibitem{jain89}
J. K. Jain,
Phys. Rev. Lett. {\bf 63}, 199 (1989)
\bibitem{jain00}
J. K. Jain,
Physics Today {\bf 53}, 39 (2000);
{\it The Composite Fermion,\/}
http://www.phys.psu.edu/~jain/cf.html
\bibitem{jainbook}
J. K. Jain, {\it Composite Fermions\/} 
(Cambridge University Press, Cambridge, England, 2007).
\bibitem{yi98}
H. Yi and H. A. Fertig, 
Phys. Rev. B {\bf 58}, 4019 (1998).
\bibitem{nare01}
R. Narevich, G. Murthy, and H. A. Fertig, 
Phys. Rev. B {\bf 64}, 245326 (2001).
\bibitem{chan05}
C-C. Chang, G. S. Jeon, and J. K. Jain, 
Phys. Rev. Lett. {\bf 94}, 016809 (2005).
\bibitem{chan06}
C-C. Chang, C. T\"{o}ke, G. S. Jeon, and J. K. Jain,
Phys. Rev. B {\bf 73}, 155323 (2006).
\bibitem{zhu10.2}
H. Zhu, Y. P. Chen, P. Jiang, L. W. Engel, D. C. Tsui, L. N. Pfeiffer, and K. W. West,
Phys. Rev. Lett. {\bf 105}, 126803 (2010).
\bibitem{zhu10.1}
H. Zhu, G. Sambandamurthy, Y. P. Chen, P. Jiang,
L. W. Engel, D. C. Tsui, L. N. Pfeiffer, and K. W. West,
Phys. Rev. Lett. {\bf 104}, 226801 (2010).
\bibitem{yl10}
C. Yannouleas and U. Landman,
Phys. Rev. A {\bf 81}, 023609 (2010).
\bibitem{yl02}
C. Yannouleas and U. Landman,
Phys. Rev. B {\bf 66}, 115315 (2002).
\bibitem{yl03}
C. Yannouleas and U. Landman,
Phys. Rev. B {\bf 68}, 035326 (2003).
\bibitem{yl04}
C. Yannouleas and U. Landman,
Phys. Rev. B {\bf 70}, 235319 (2004).
\bibitem{yl07}
C. Yannouleas and U. Landman,
Rep. Prog. Phys. {\bf 70}, 2067 (2007).
\bibitem{note31}
The effects of disorder (pinning) at zero magnetic field in stabilizing the Wigner crystal against the 
liquid phase, resulting in a shifting of the crossover between these phases to higher electron
densities, have been discussed in T. Chui and B. Tanatar, Phys. Rev. Lett. {\bf 74}, 458 (1995).
Impurity pinning effects of Wigner molecules (by both attractive and repulsive impurities) 
have been studied theoretically in the context of electrons confined in 2D semiconductor quantum dots;
for magnetic-field-free conditions, as well as for a finite magnetic field
(for up to eight electrons), see C. Yannouleas and U. Landman, 
Phys. Rev. B {\bf 61}, 15895 (2000), and for a study with an applied magnetic field (for $N=3$ and 
$N=4$ electrons), see B. Szafran and F. M. Peeters, Europhys. Lett. {\bf 66}, 701 (2004).
\bibitem{jain95}
J. K. Jain and T. Kawamura, 
Europhys. Lett. {\bf 29}, 321 (1995).
\bibitem{lyl06}
Yuesong Li, C. Yannouleas, and U. Landman,
Phys. Rev. B {\bf 73}, 075301 (2006).
\bibitem{taka86}
K. Takano and A. Isihara,
Phys. Rev. B {\bf 34}, 1399 (1986).
\bibitem{yang02}
X. Wan, K. Yang, and E. H. Rezayi,
Phys. Rev. Lett. {\bf 88}, 056802 (2002).
\bibitem{wexl03}
O. Ciftja and C. Wexler,
Phys. Rev. B {\bf 67}, 075304 (2003).
\bibitem{yang03}
X. Wan, E. H. Rezayi, and K. Yang,
Phys. Rev. B {\bf 68}, 125307 (2003).
\bibitem{jola09}
S. Jolad and J. K. Jain,
Phys. Rev. Lett. {\bf 102}, 116801 (2009).
\bibitem{jola10}
S. Jolad, D. Sen, and J. K. Jain,
Phys. Rev. B {\bf 82}, 075315 (2010).
\bibitem{girv83}
S. M. Girvin and T. Jach, 
Phys. Rev. B {\bf 28}, 4506 (1983).
\bibitem{note1}
Vibrational excitations of a similar form, i.e., 
\begin{equation}
\tilde{Q}_\lambda=\sum_{i=1}^N z_i^\lambda \nonumber
\end{equation}
(and certain other variants), have been used earlier than Ref.\ \onlinecite{yl10} to 
approximate {\it part\/} of the LLL spectra. Such earlier endeavors provided valuable 
insights, but overall they remained inconclusive; for electrons over the maximum density 
droplet [with magic $L_m=L_0$], see M. Stone, H. W. Wyld, and R. L. 
Schult, Phys. Rev. B {\bf 45}, 14 156 (1992) and J. H. Oaknin, L. Mart\'{i}n-Moreno, J. J. 
Palacios, and C. Tejedor, Phys. Rev. Lett. {\bf 74}, 5120 (1995); for electrons over the 
$\nu=1/3$  Jastrow-Laughlin trial function [with magic $L_m=3 L_0$], see J. J. 
Palacios and A. H. MacDonald, Phys. Rev. Lett. {\bf 76}, 118 (1996); and for bosons in the 
range $0 \leq L \leq N$, see B. R. Mottelson, Phys. Rev. Lett. {\bf 83}, 2695 (1999) and Th. 
Papenbrock and G.F. Bertsch, Phys. Rev. A {\bf 63}, 023616 (2001). The advantage of 
$Q_\lambda$ (compared to $\tilde{Q}_\lambda$) is that it is translationally invariant, 
\cite{yl10} a property shared with $\Phi^{\text{REM}}_{\cal L}$.
\bibitem{note2}
The RVEM diagonalization provides a unified molecular treatment that can also be applied
to an assembly of $N$ bosons in the LLL, as shown in Ref.\ \onlinecite{yl10}.
\bibitem{note3}
For higher angular momenta, the cusp states progressively correspond exlusively to the (1,5) 
ring configuration, which classically is the most stable one; see Ref.\ \onlinecite{yl03}.
\bibitem{beda94}
V. M. Bedanov and F. M. Peeters,
Phys. Rev. B {\bf 49}, 2667 (1994).
\bibitem{kong03}
M. Kong, B. Partoens, and F. M. Peeters,
Phys. Rev. E {\bf 67}, 021608 (2003).
\bibitem{bao95}
W. Y. Ruan, Y. Y. Liu, C. G. Bao, and Z. Q. Zhang,
Phys. Rev. B {\bf 51}, 7942 (1995).
\bibitem{seki96}
T. Seki, Y. Kuramoto, and T. Nishino, 
J. Phys. Soc. Jpn. {\bf 65}, 3945 (1996).
\bibitem{maks96}
P. A. Maksym,
Phys. Rev. B {\bf 53}, 10871 (1996).
\bibitem{tave03}
M. B. Tavernier, E. Anisimovas, F. M. Peeters, B. Szafran, J. Adamowski, and S. Bednarek,
Phys. Rev. B {\bf 68}, 205305 (2003).
\bibitem{tsip01}
E. V. Tsiper and V. J. Goldman,
Phys. Rev. B {\bf 64}, 165311 (2001).
\bibitem{note4}
Deviations of the same nature (i.e., overestimation of the vibrational components) 
between the EXD and Laughlin-state radial EDs have been reported for the whole range 
$5 \leq N \leq 12$ of LLL electrons (see Ref.\ \onlinecite{tsip01}); these deviations 
increase slowly with increasing $N$.
\bibitem{note6}
For a $(0,N)$ configuration, the magic angular momenta $L_m$ are given by $L_m=L_0+kN$,
$k=0,1,2, \ldots$ [for $L_0$ see Eq.\ (\ref{l0})]; i.e., for $N=6$: 
$L_m=15$, 21, 27, 33, 39, 45, 51, $\ldots$.
Similarly, for a $(1,N-1)$ configuration, $L_m=L_0+k(N-1)$ ($k=0,1,2, \ldots$), giving for
$N=6$: $L_m=15$, 20, 25, $\ldots$, 40, 45, 50, $\ldots$.  
\bibitem{note7}
Any nonrotating (static) electron-molecule state is a wave packet, and thus it is
expandable in a complete basis of rotating electron molecule states with good total
angular momenta. Therefore, given the wave function of the nonrotating EM, a state with
good total angular momentum can be obtained from it via an appropriate projection. This
constitutes the direct projection method developed by us in the context of ``restoration of
broken symmetries'' (see Refs.\ \onlinecite{yl02,yl03,yl07}). The term ``reverse projection'' 
is used to indicate the ``inverse'' operation to the direct projection, namely, the method of 
generating the wave function of a nonrotating EM out of the states (with good angular
momentum) of a rotating and vibrating electron molecule. 
\bibitem{goer04}
M. O. Goerbig, P. Lederer, and C. Morais Smith,
Phys. Rev. Lett. {\bf 93}, 216802 (2004).
\bibitem{goer04.2}
M. O. Goerbig, P. Lederer, and C. Morais Smith,
Phys. Rev. B {\bf 69}, 115327 (2004).
\bibitem{note5}
In addition to the second-half (starting with ``Both liquid and solid phases ...'') of the 
introductory paragraph in Ref.\ \onlinecite{goer04}, the inapplicability of the Hamiltonian CF 
approach to filling-factor ranges around integer values can be explicitly traced to Ref.\ 
\onlinecite{goer04.2}; see in particular the sentence ``We further require 
that the partial filling factor of the last level $\bar{\nu}=\nu-[\nu]$ is different from zero 
because at integral fillings, the only possible low-energy excitations are inter-LL excitations,
which cost an energy of order $\hbar \omega_c$ ...'' in the first paragraph of Sec. II therein.
\bibitem{note24}
An attempt to formulate a variant of a CF Wigner crystal, referred to simply as CFC, for a 
finite number of electrons $N$ was presented in Refs.\ \onlinecite{chan05,chan06}. This CFC 
formulation, however, is explicitly inapplicable to the $\nu=1/3$ neighborhood;
see Sec.\ 15.2 (p. 450) in Ref.\ \onlinecite{jainbook}. 
\bibitem{jeon07}
G. S. Jeon, C. C. Chang, and J. K. Jain,
Eur. Phys. J. B {\bf 55}, 271 (2007). 
\bibitem{jain98}
J. K. Jain and R. K. Kamilla,
in {\it Composite Fermions: A Unified View of the Quantum Hall Regime,\/}
Edited by O. Heinonen (World Scientific, Singapore, 1998).
\end{thebibliography}
\end{document}